# Ultrafast THz probing of nonlocal orbital current in transverse multilayer metallic heterostructures[#]

Sandeep Kumar and Sunil Kumar[*]
*Femtosecond Spectroscopy and Nonlinear Photonics Laboratory,
Department of Physics, Indian Institute of Technology Delhi, New Delhi 110016, India*
*Email: [kumarsunil@physics.iitd.ac.in](kumarsunil@physics.iitd.ac.in)

THz generation from femtosecond photoexcited spintronic heterostructures has recently become a versatile tool for investigating ultrafast spin-transport and transient charge-current in a non-contact and non-invasive manner. The same from the orbital effects is still in the primitive stage. Here, we experimentally demonstrate orbital-to-charge current conversion in metallic heterostructures, consisting of a ferromagnetic layer adjacent to either a light or a heavy metal layer, through detection of the emitted THz pulses. Temperature-dependent experiments help to disentangle the orbital and spin components that are manifested in the respective Hall-conductivities, contributing to THz emission. NiFe/Nb shows the strongest inverse orbital Hall effect with an experimentally extracted value of effective intrinsic Hall-conductivity, $(\sigma_{SOH}^{int})^{eff} \sim 195\, \Omega^{-1} cm^{-1}$, while CoFeB/Pt shows maximum contribution from the inverse spin Hall effect. In addition, we observe nearly ten-fold enhancement in the THz emission due to pronounced orbital-transport in W-insertion heavy metal layer in CoFeB/W/Ta heterostructure as compared to CoFeB/Ta bilayer counterpart.

---

# 1. INTRODUCTION

Efficient generation and detection of spin current are the key requirements in spintronic devices for their different potential applications.[1,2] These include, the spin-orbit torque (SOT), spin-pumping, magnetic memories, excitation of magnons, manipulating the magnetic damping, etc. Mainly, the spin Hall effect[3,4] (SHE) and Rashba-Edelstein effect[5,6] (REE), governed by the spin angular momentum (S) transfer, have been invoked commonly in the generation of a spin current from a charge current. It has become clear from a few recent studies that in certain solids,[7-9] the transport of electron's orbital angular momentum (L) and hence the associated magnetic moment, is also responsible for various interesting phenomena in the emerging field of orbitronics. Orbital Hall effect (OHE), which was conceived[10,11] just after the SHE,[3] has often been neglected due to orbital quenching in the periodic solids.[12,13] Bearing many similarities with the SHE, in the OHE, a transverse flow of orbital angular momentum occurs in response to a longitudinally applied electric field. In fact, fundamentally, SHE has been proposed to originate from OHE only.[14-16] Therefore, it has helped in resolving the sign and magnitude of the reported values of spin Hall conductivities for certain materials. A few theoretical and experimental studies in the recent literature[13,14,17] have indicated a gigantic OHE in the light as well as several heavy metals and therefore, it has necessitated more careful investigations of SHE-based phenomena and devising new schemes to disentangle the OHE.

As indicated above, OHE is a fundamental phenomenon, which can be observed in a variety of materials, including transition metals[13-15,17-20], semiconductors[11], two-dimensional materials[21-23], etc.[24] Unlike the SHE, whose strength greatly depends on the spin-orbit coupling (SOC) in the material, OHE, on the other hand, can be found even in the light materials, having very weak SOC[13,20]. After a few theoretical reports[13,14,17] on the large orbital Hall conductivity ($\sigma_{OH}$) in some strong as well as weak SOC-type materials, detailed experimental demonstrations are required to obtain further insights for harnessing the same in practical applications. For instance, a high value of the Hall conductivity is always advantageous as it helps in the enhancement of spin-orbit torque (SOT), which is technologically relevant to memory applications. Conventionally, in SHE induced torque (SHT), the spin angular momentum transfer exerts a torque directly on the local magnetization of the material. However, due to the lack of direct exchange coupling[7] of orbital angular momentum with the local magnetization, a similar direct realization of the OHE induced torque or the orbital Hall torque (OHT) was lacking. This pertinent issue has found a regenerated interest among researchers[7,8,17,19] to make OHT based applications viable[9,25] by devising novel orbital-spin (L-S) conversion schemes and suitable material combinations.[19,26] It follows from the magnetization manipulation through the exerted net torque, dominated by either SHT or OHT, and acts as a key in distinguishing the orbital character indirectly from the pure spin transport, yet not in an unambiguous manner. Y.G. Choi *et al.*,[16] have used magneto-optical Kerr effect (MOKE) in addition to the orbital torque measurement technique for direct detection of orbital magnetic moment accumulation created by the charge current flow in a light metal, Ti. A non-contact method is always promising to non-invasively measure the spin and orbital transport in materials.

Like the toque method to detect the charge to spin or orbital conversion, the inverse of the SHE (ISHE) and REE (IREE) are routinely used for the detection of spin transport through the spin-charge conversion, where the spin source can be one from either spin pumping or spin Seebeck current or optical excitation, etc. The Onsager reciprocity[27] allows an interconversion between the orbital and charge currents, where, similar to the ISHE and IREE for the spin counterpart, here, inverse orbital Hall effect (IOHE) and inverse orbital Rashba-Edelstein effect[25] (IOREE) are both in play. In several studies[28-37] in the last decade or so, ISHE and IREE have been utilized in the THz electromagnetic pulse generation from ultrafast photoexcited magnetic/nonmagnetic (NM) multilayer systems. Consequently, the scheme, in conjunction with the multilayers, is not only recognized as a source of effective THz radiation,[38] but also a highly sensitive contactless optical probing tool[39-46] for the detection and control of the ultrafast processes at femtosecond time scale, spin-charge conversion mechanisms, demagnetization dynamics and transport, interfacial properties, etc. For the case of the ISHE based spintronic THz emitters, spin current from the ferromagnetic (FM) or antiferromagnetic layer is injected into the NM layer, where it gets converted to charge current. Therefore, heavy metal layer with large SOC is desirous for efficient THz generation from a bilayer[47]. Similar effects are envisaged to exist for the orbital counterpart too.[48,49] For the THz emission utilizing the orbital transport properties, i.e., orbital-charge conversion through IOHE, material candidates capable of generating an orbital current ($J_L$) and subsequently converting the same into charge current, are desired.

In the current work, existence of nonlocal orbital transport is experimentally detected through IOHE mediated efficient THz emission from femtosecond NIR (near-infrared) pulse excited bi- and tri-layer metallic heterostructures using temperature-dependent time-domain spectroscopy that has not been reported hitherto. Since the orbital degree of angular momentum is strongly correlated with the crystal field potential, therefore, the temperature-dependency of the phonon scattering would severely affect the OHE and the related phenomena, microscopically. The specifically chosen heterostructures, in this work, consist of FM and NM material combinations, where the choice of the FM is from either CoFeB or NiFe, whereas the NM is from both the light metal (Nb) as well as the heavy metals (Pt, Ta, and W). While the THz emission from CoFeB/Pt, CoFeB/Ta, CoFeB/W and Fe/Ta bilayers is shown to originate principally from the ISHE in the heavy metal layers therein, the same from NiFe/Nb arises primarily via the IOHE in the light metal layer of Nb. In the prior, strong ultrafast photoinduced spin current is generated in the FM layer to be further injected into the NM heavy metal layer whereas in the latter case, efficient spin-orbital conversion within NiFe layer facilitates strong ultrafast orbital current injection into the Nb layer. The temperature-dependence of the THz amplitude vis-a-vis the Hall conductivities are used to distinguish spin-to-charge and orbital-to-charge signatures in the NM layers. The wide-range temperature-dependent THz results and analysis also help to distinguish dominating extrinsic and intrinsic contributions to IOHE in different resistivity regions. For the observation of IOHE mediated THz emission from structures consisting of heavy metal layer, we fabricated a tri-layer system of CoFeB/W/Ta and measured the temperature-dependent THz amplitude and the Hall conductivities. The THz emission from such a tri-layer, having the W-insertion layer, interfaced with another heavy metal layer of same sign of the spin Hall angle and placed side-by-side, is nearly one-order stronger than the CoFeB/Ta bilayer counterpart. Such an enhancement in the THz emission is associated with efficient spin to orbital conversion due to strong SOC and long diffusion length of the orbital current in the W-insertion layer. The next section provides our results and detailed discussion on them sequentially. First, the case of NiFe/Nb bilayer is taken up, followed by the study on CoFeB/NM (Pt,Ta) bilayers and, finally, the trilayer of CoFeB/W/Ta. For completeness, the results on CoFeB/W bilayer, also particular comparisons with CoFeB/Ta and CoFeB/W/Ta heterostructures, are described in the Supplementary Information document. All these heterostructures are grown on quartz substrates by using UHV RF sputtering. A nearly optimized thickness and phase of the individual layers in the heterostructures has been used. The W and Ta layers are in their α-phase. For all the THz results reported here, femtosecond pulses having time-duration of ~50fs, central wavelength of 800nm, pulse energy (fluence) of ~35μJ (0.5mJ/cm$^2$) were used. However, results with the varying pump fluence on different samples are shown in the Supplementary Section S11. Complete details about the experimental arrangements in the THz setup, material synthesis and characterization, are provided in the Supplementary Information. Briefly, they are mentioned in the Methods section also.

## 2. RESULTS AND DISCUSSION

### 2.1 NiFe/Nb: Probing inverse orbital Hall effect in light metal through generation of THz pulses

Figure 1(a) schematically illustrates the emission of THz pulses from the NiFe/Nb bilayer following optical excitation by linearly polarized femtosecond NIR pulses whereas, the same for bare NiFe sample is presented in the Supplementary Section S12. The thicknesses of the layers were kept at 5 nm for NiFe and 10 nm for Nb layer. The full THz bandwidth of the signals from NiFe/Nb sample as obtained by fast Fourier transform is shown in the Supplementary Section S6. A constant external magnetic field (B), having a value just above the saturation (~200 Oe), is applied along the y-direction. A few consistency checks were recorded, to initially validate the origin for the generation of THz pulses, for four geometries of the direction of the optical excitation and the magnetic field, as presented in Figs. 1(b) and 1(c), respectively. NiFe is a popular FM material for spintronic applications. Dominant presence (≥90%) of Ni (light element) in NiFe makes it possess a large and positive value of spin-orbit correlation factor,[8] $\langle L.S \rangle = \eta > 0$. As shown in Fig. 1(d), in positive spin-orbit correlation materials, transverse orbital and spin Hall effects are induced in response to the flow of a longitudinal charge current, $J_C$ such that the polarization direction of the accumulated orbital and spin magnetic moments is the same.[8] An optimal material composition and growth of NiFe enriched with Ni can provide a better value of η as compared to that in Ni.[50] Moreover, a transient change in the spin-orbit coupling in Ni, triggered by the high energy ultrafast excitation, has been seen to enhance the η value significantly.[51]

Following the optical pulse excitation of NiFe/Nb (Fig. 1a), ultrafast demagnetization[52] in the NiFe layer stimulates flow of a spin current with density $J_S$. Due to a large positive value of η in the NiFe, a fraction of the ultrafast spin current is converted into an ultrafast orbital current ($J_L$) of same polarity through the L-S conversion, given[8,48] as, $J_L = \eta_{L-S} \cdot J_S$. Therefore, an ultrafast optically induced orbital current sets in,[48,53] which possess similar symmetry properties to the spin current but can exhibit relatively different transport dynamics.[9,54,55] Furthermore, as the ultrafast excitation of spin and orbital magnetization has been reported[51,56-58] to exhibit a similar evolution, the emergence of orbital current can also be comprehended through the analogy with the already established spin current formation.[53] Consequently, as indicated in Fig. 1(a), both the spin and orbital currents through their interconversion within the NiFe layer are now launched into the adjacent Nb layer. A very weak negative SOC strength and about an order larger orbital Hall conductivity than the spin Hall conductivity ($\sigma_{OH}$) in Nb make it a suitable candidate[59,60] for realizing orbital transport phenomena. In case of heavy metals, although the value of $\sigma_{OH}$ is typically much larger than the $\sigma_{SH}$,[14] however, OHE is greatly suppressed by the inherently present SOC owing to the identical macroscopic geometries between OHE and SHE in response to the applied electrical current in steady operations.[20] Therefore, to overcome the effect of spin transport in the realization of OHE, selection of an appropriate light element material is regarded as one of the key solutions.

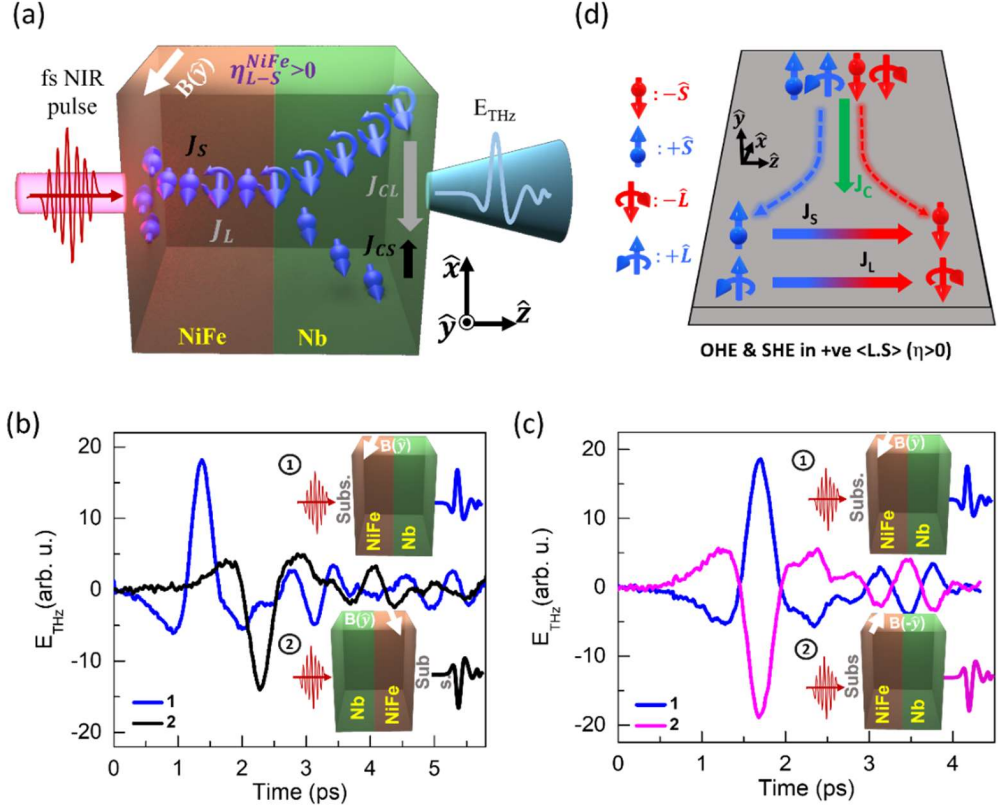

**Fig. 1. Orbital-to-charge current conversion mediated THz emission from NiFe/Nb bilayer.** (a) Schematic illustration of ultrafast optically induced spin current $J_S$ and orbital current $J_L$ in the NiFe and their injection into the Nb layer producing ultrafast charge current and hence emission of THz pulses. The positive value of $\eta_{NiFe}$ represents the spin-to-orbital current conversion efficiency in NiFe. $J_{CS}$ and $J_{CL}$ represent charge currents converted from $J_S$ and $J_L$ through the ISHE and IOHE, respectively. $B$ represents the externally applied magnetic field. (b) Temporal profiles of THz signal obtained from NiFe/Nb under four experimental geometries with the direction of the optical excitation and the magnetic field: (b) 1. substrate side excitation; 2. Nb film side excitation, while keeping B fixed along $\hat{y}$-direction, (c) 1. $B(+\hat{y})$; 2. $B(-\hat{y})$, while keeping the optical excitation from the substrate side. (d) Schematic of OHE and SHE in a prototype material. A charge current, $J_C$ in the $-\hat{x}$-direction induces accumulation of spin (S) and orbital (L) angular momenta in the transverse $\hat{z}$-direction. For positive spin-orbit correlation, $\eta > 0$, both the spin and orbital polarizations are parallel along either $+\hat{y}$-direction or $-\hat{y}$-direction.

The spin and orbital currents injected into the Nb layer convert to respective in-plane transient charge currents. Charge current, $J_{CS}$ is produced via ISHE and $J_{CL}$ is produced via IOHE. The magnitude and polarity of the emitted THz radiation is finally dependent on the net transient charge current in the Nb layer. Typically, the orbital diffusion length in Nb is comparable to that of spin diffusion length (see Supplementary Section S13). Due to the negligible SOC and correspondingly smaller spin Hall angle in Nb, the ISHE signal is nearly quenched (depicted by small black vector within the Nb layer in Fig. 1(a)) as against the IOHE. The net charge current in the Nb layer is the vectorial sum of the two charge currents, i.e., $J_C = J_{C-ISH} + J_{C-IOHE}$, and it can be further expressed as

$$J_C = \theta_{SH} \cdot J_s + \theta_{OH} \cdot J_L = \theta_{SH}^{Nb} \cdot J_s + \theta_{OH}^{Nb} \cdot \eta_{L-S}^{NiFe} \cdot J_s \qquad (1)$$

Here, $\theta_{SH}$ and $\theta_{OH}$ are the spin and orbital Hall angles of Nb, which are negative and positive, respectively.[15,18] Since the spin-orbit correlation factor, $\eta_{L-S}^{NiFe}$ is positive for the NiFe, therefore, the charge current contribution from the individual components would have opposite signs, as represented in Fig. 1(a). The net charge current is, therefore, the difference of the two, and the dominant one among them would control the amplitude and polarity of the emitted THz radiation. If the THz signal from the NiFe/Nb bilayer structure is a result of the inverse spin Hall effect (ISHE), its polarity would be expected to be the same as that of the Fe/Ta bilayer but opposite to that of the CoFeB/Pt bilayer structures. This distinction arises because Nb and Ta both have the negative sign of $\theta_{SH}^{Ta}$, while Pt has a positive sign of $\theta_{SH}^{Pt}$. However, the observed polarity of the THz signals in the NiFe/Nb bilayer structure is opposite to that of Fe/Ta bilayer structure but coincides with the CoFeB/Pt bilayer structure, as schematically depicted in Figs. 1(a), 3(a), and 3(b), respectively. In fact, the polarity of the THz signals emitted from NiFe/Nb bilayer structure aligns with the sign of $\theta_{OH}^{Nb} \cdot \eta_{L-S}^{NiFe}$ implying that the resultant THz emission from NiFe/Nb takes place via IOHE, mainly. A few consistency checks are carried out in Figs. 1(b) and 1(c) to reveal the magnetic origin of the THz signal. The THz signal polarity is inverted by reversing either the direction of the optical excitation while keeping the magnetic field unchanged or the direction of the external magnetic field while keeping the direction of optical excitation unchanged. In the first case, the THz sign reversal is due to the change in the direction of the spin current and consequently the flow of orbital current via spin to orbit conversion in NiFe, whereas, in the second case, it is due to the change in the spin current polarization and subsequently the polarization of orbital current via spin-orbit conversion as both are constrained by positive $\eta_{L-S}^{NiFe}$.

Temperature-dependent studies[61-63] have been proven to be important in identifying various scattering mechanisms in the materials and subsequent determination of respective Hall conductivities. To validate the IOHE origin of the THz emission from NiFe/Nb bilayer, we have performed temperature-dependent experiments by monitoring the changes in the peak-to-peak amplitude of the THz signal as a function of the sample temperature (T) from 10 K to 300 K. The temperature sensitivity of the phonon scattering would have a significant impact on the OHE and associated phenomena because the orbital degree of angular momentum is tightly coupled with the crystal field potential.[13,20]. The result is presented in Fig. 2(a), where the error bars represent the largest absolute deviation from the mean of three THz signal outputs at each temperature. As indicated in the inset of the figure, the sample is optically excited from the substrate side and the same orientation for sample excitation is followed in all the results presented below in the paper. The substrate side optical excitation geometry does not create any temperature dependent contributions (see Supplementary Section S15). Since, the sign and magnitude of the emitted THz pulse is directly related with the $\sigma_{SH}$ or $\sigma_{OH}$, hence, by looking at the THz signal's amplitude and phase, a qualitative information about the $\sigma_{SH}$ or $\sigma_{OH}$ can be obtained.[32] For exactly determining the dominating role of either the $\sigma_{SH}$ or $\sigma_{OH}$, detailed analysis is presented below.

By employing the four-point van der Pauw method, electrical longitudinal resistivity ($\rho$) was also determined for the Nb and NiFe/Nb films in the entire experimental temperature range (see inset of Fig. 2(a)). The resistivity information, together with the THz amplitude data in Fig. 2(a), are used to obtain the results presented in Fig. 2(b), where we have plotted the behavior of extracted effective spin-orbital Hall resistivity ($\rho_{SOH}^{eff}$) with respect to the squared longitudinal resistivity ($\rho_{NM}^2$) of NM = Nb. See Supplementary Section S10 for details to obtain $\rho_{SOH}^{eff}$. In the light of a temperature scaling relation[61-64] for the spin Hall resistivity ($\rho_{SH}$),

$$\rho_{SH}(T) = \sigma_{SH}^{int.} \cdot \rho_{NM}^2(T) + \sigma_{SJ} \cdot \rho_{0,NM}^2 + \alpha_{ss} \cdot \rho_{0,NM} \quad (2)$$

Here, $\sigma_{SH}^{int.}$, $\sigma_{SJ}$, $\rho_{0,NM}$, $\alpha_{ss}$ are intrinsic spin Hall conductivity, side-jump spin Hall conductivity, residual resistivity, and skew scattering angle of the NM layer, respectively. The second and the third terms on the right-hand side of Eq. (2) represent the extrinsic contributions to scattering. Because of the correspondence between the SHE and OHE, a similar phenomenological equation for the orbital Hall resistivity ($\rho_{OH}$) can be constructed by accounting for the intrinsic and extrinsic orbital scattering processes.[8,15,20] Hence, temperature-dependence of $\rho_{OH}$ can be expressed as,

$$\rho_{OH}(T) = \eta_{L-S} \cdot \sigma_{OH}^{int.} \cdot \rho_{NM}^2(T) + Ex \quad (3)$$

On the right side of the above equation, the first term, consisting of the intrinsic orbital Hall conductivity, $\sigma_{OH}^{int}$, represents the intrinsic scattering, while the extrinsic contributions to orbital Hall resistivity are represented by the second term, $Ex$. The factor, $\eta_{L-S}$, takes care of the spin-orbit interconversion arising due to SOC. The extrinsic contribution to the spin or the orbital Hall resistivity in the above equations, is usually neglected for pure materials[63,65] while, it can be significantly high for materials with high impurity concentration. For capturing the temperature-dependence of the effective spin-orbital Hall resistivity, Eqs. (2) and (3) can be combined and $\rho_{SOH}^{eff}$ is expressed at each temperature $T$ as following,

$$\rho_{SOH}^{eff}(T) = (\sigma_{SH}^{int} + \eta_{L-S} \cdot \sigma_{OH}^{int}) \cdot \rho_{NM}^2(T) = (\sigma_{SOH}^{int})_{NM}^{eff} \cdot \rho_{NM}^2(T) \quad (4)$$

In the above, we have used $(\sigma_{SOH}^{int})^{eff} = (\sigma_{SH}^{int} + \eta_{L-S} \cdot \sigma_{OH}^{int})$ to represent the effective intrinsic spin-orbital Hall conductivity and any weak temperature variations in the extrinsic terms in both the spin and the orbital Hall resistivities have been ignored. As shown in Section S10 of the Supplementary Information, the effective spin-orbital Hall resistivity at given temperature $T$ is related to the amplitude of the THz signal,[61,64] and it is given by the relation,

$$\rho_{SOH}^{eff}(T) \sim E_{THz}(T) \left(\frac{\rho_{xx}^{NM}}{\rho_{FM/NM}}\right) \left(\frac{d}{\lambda_{LS}}\right) \frac{1}{e \cdot J_S} \quad (5),$$

where, $\lambda_{LS}$ is the effective spin-orbit diffusion length[17] and $J_S$ still represents the maximum instantaneous spin current produced in the FM layer by the ultrafast optical excitation. The extracted data for $\rho_{SOH}^{eff}$ vs $\rho_{NM}^2$ at each temperature, are presented in Fig. 2(b). From the behavior of the data, two distinct resistivity regions, above and below $\rho_{Nb}^2 \sim 0.5 \times 10^3$ $(\mu\Omega \cdot cm)^2$, are evident. Such an observation with respect to the squared longitudinal resistivity has been found helpful in probing intrinsic and extrinsic contribution dominated resistivity regions for AHE[66-69] and SHE.[63] However, the same for the case of OHE, though anticipated[17], but has not been reported experimentally so far. According to Eqs. (4) and (5), linear fits to the data in Fig. 2(b) yield $(\sigma_{SOH}^{int})_{Nb}^{eff} \sim + 195$ $(\hbar/e)$ $\Omega^{-1} cm^{-1}$ in the low-resistivity region and $(\sigma_{SOH}^{int})_{Nb}^{eff} \sim + 1210$ $(\hbar/e)$ $\Omega^{-1} cm^{-1}$ in the high resistivity region. From the theoretically known values of different intrinsic parameters for NiFe/Nb sample, i.e., $\sigma_{SH}^{int}((\hbar/e)\,\Omega^{-1}cm^{-1}) \cong -100^{14,18}$, $\eta_{L-S} \cong 0.045^8$, and $\sigma_{OH}^{int}((\hbar/e)\,\Omega^{-1}cm^{-1}) \cong 6000^{18}$, we obtain, $(\sigma_{SOH}^{int})_{Nb}^{eff} = (\sigma_{SH}^{int} + \eta_{L-S} \cdot \sigma_{OH}^{int}) \sim +170\ (\hbar/e)\,\Omega^{-1}cm^{-1}$, a value matching with our experimental value. The excellent agreement between the two values clearly indicates dominance of orbital transport with majorly intrinsic contribution to OHE in the high resistivity region of the NiFe layer. Therefore, we conclude from here that dominating orbital current in the NiFe layer governs the THz emission from NiFe/Nb via intrinsic IOHE in the light metal Nb layer. The much higher value of $(\sigma_{SOH}^{int})^{eff} \sim +1210\ (\hbar/e)\,\Omega^{-1}cm^{-1}$ in the low resistivity region must originate from extrinsic reasons, analogous to what is known for AHE[66-69] and SHE.[63] With the information at hand that for Nb, $\sigma_{SH}^{int}$ is negative and $\eta_{L-S}$ is positive, a positive value of $(\sigma_{SOH}^{int})_{Nb}^{eff}$ from the fit clearly reveals that $\eta_{L-S} \cdot \sigma_{OH}^{int}$ for Nb is a large positive value and dominates over $\sigma_{SH}^{int}$. In fact, theoretical predictions of a gigantic +ve value of $\sigma_{OH}$ over a small -ve value of $\sigma_{SH}$ in Nb[70], are well aligned with our experimental observations.

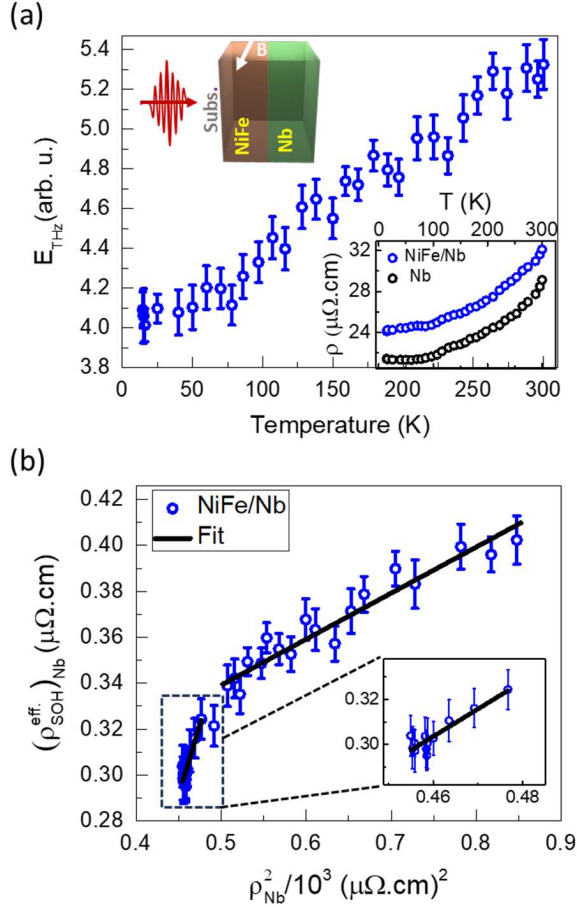

**Fig. 2. Temperature-dependent THz emission from NiFe/Nb (Nb = light metal).** (a) Peak-to-peak value of THz amplitude as a function of varying sample temperature. The error bars at each temperature correspond to the largest absolute deviation of the peak-to-peak THz amplitude from the mean of three measurements. Inset: The optical excitation from the substrate side; temperature-dependent resistivity (ρ) variations of NiFe/Nb and Nb samples measured by the four-point van der Pauw method. (b) Effective spin-orbital Hall resistivity as a function of the squared longitudinal resistivity for the Nb film. Inset: Zoomed-in view of the data below $\rho_{Nb}^2 \sim 0.5 \times 10^3\ (\mu\Omega.cm)^2$. The solid lines are fit to the data using Eq. (4).

**2.2 FM(CoFeB,Fe)/NM(Pt,Ta): ISHE mediated THz emission from heavy metal layers based FM/NM heterostructures**

Experimental results are now presented and discussed on Fe/Ta and CoFeB/Pt, in conjunction with those on CoFeB/Ta, presented elsewhere,[61] and re-examined again as presented in the Supplementary Section S8. These bilayer heterostructures contain a heavy metal layer, either Pt or Ta, and are chosen selectively for their opposite sign of the spin Hall angle.[32,34] Moreover, CoFeB and Fe are selected because of their negligible[8,48] spin-orbit correlation factor, i.e., $\eta_{L-S}^{CoFeB}$ and $\eta_{L-S}^{Fe} \sim 0$ as compared to Ni or NiFe presented earlier. By these choices, we initially ensure that there is negligible fractional conversion of the spin current into orbital current within the FM layer. Consequently, from the femtosecond pulse excited FM/NM bilayers, THz generation takes place via ISHE only. For the CoFeB/Pt with the Pt layer having positive spin Hall angle ($\theta_{SH}^{Pt} >$ 0), the experimental configuration is depicted in Fig. 3(a), where the directions of the external magnetic field, spin current, charge currents and the THz signal polarities are indicated. Owing to the opposite sign of the spin Hall angle in Ta ($\theta_{SH}^{Ta} < 0$), emission of opposite polarity THz signal from Fe/Ta, under the same experimental configuration, is shown in Fig. 3(b). In both cases, the THz signal polarity dependence on the experimental configuration, including the direction of the external magnetic field and the optical excitation, are found to be as expected from the ISHE mediated THz emission.[29,32] Moreover, ultrafast demagnetization mechanism is mainly responsible for the THz emission from the bare FM layer (see Supplementary Section S12). Variations of the peak-to-peak THz signal amplitude as a function of the sample temperature for the CoFeB/Pt and Fe/Ta bilayers, are presented in Figs. 3(c) and 3(d), respectively. The temperature-dependent longitudinal resistivities for both the samples is provided in the Supplementary Section S7. From the experimentally measured THz amplitude and resistivities, we have derived the effective spin-orbital Hall resistivity, $\rho_{SOH}^{eff}$ at each temperature using the procedure discussed in Section S10 of Supplementary Information, and the results for the same as a function of squared longitudinal resistivity, $\rho_{NM}^2$ ($NM = Pt, Ta$) are presented in Figs. 3(e) and 3(f) for CoFeB/Pt and Fe/Ta, respectively. In the next paragraph, we establish sole ISHE origin of the THz signal generation through the dominance of the extracted respective spin Hall conductivities in CoFeB/Pt and Fe/Ta bilayers.

We fit the results in Figs. 3(e) and 3(f) using Eq. (4) to obtain $(\sigma_{SOH}^{int})^{eff}$ from the slope of the linear fit. Firstly, in the case of Fe/Ta, a negative slope value, i.e., $(\sigma_{SOH}^{int})^{eff} = (\sigma_{SH}^{int} + \eta_{L-S} \cdot \sigma_{OH}^{int}) < 0$, is obtained. As it is well known in the literature that for Ta, $\sigma_{SH}^{int}$ is negative, while, $\sigma_{OH}^{int}$ is positive.[14,18] Also, the spin-orbit correlation factor for Fe, $(\eta_{L-S})$, is known to be insignificantly positive.[8] Therefore, a negative value of $(\sigma_{SOH}^{int})^{eff}$ from the experiments directly implies that $\sigma_{SH}^{int} > \eta_{L-S} \cdot \sigma_{OH}^{int}$. Moreover, positive $\sigma_{OH}$ value of Ta demands the slope to be positive in Fe/Ta sample, which is not the case here. Hence, it can be concluded that the THz emission from Fe/Ta is entirely due to the spin-to-charge conversion via ISHE in Ta. On the other hand, a positive slope of the fit, i.e., $(\sigma_{SOH}^{int})^{eff} > 0$ is obtained in the case of CoFeB/Pt. Since, $\sigma_{SH}^{int}$, $\sigma_{OH}^{int}$ and $\eta_{L-S}$, all are positive for Pt,[14,18] the positive valued effective conductivity, $(\sigma_{SOH}^{int})^{eff} = (\sigma_{SH}^{int} + \eta_{L-S} \cdot \sigma_{OH}^{int})$ is as per the expectation for dominating spin Hall conductivity. Thanks to the negligible value of $\eta_{L-S}^{CoFeB}$, the contribution, $\eta_{L-S} \cdot \sigma_{OH}^{int}$ becomes far smaller than $\sigma_{SH}^{int}$, and this would results into the condition $\sigma_{SH}^{int} > \eta_{L-S} \cdot \sigma_{OH}^{int}$, same as in the case of Ta. The dominating $\sigma_{SH}^{int}$ dictates flow of majorly a spin current from CoFeB and its conversion to charge current takes place in Pt via ISHE. Therefore, it can be concluded from here that ISHE is the origin for THz pulse emission takes place in the CoFeB/Pt and Fe/Ta bilayers.

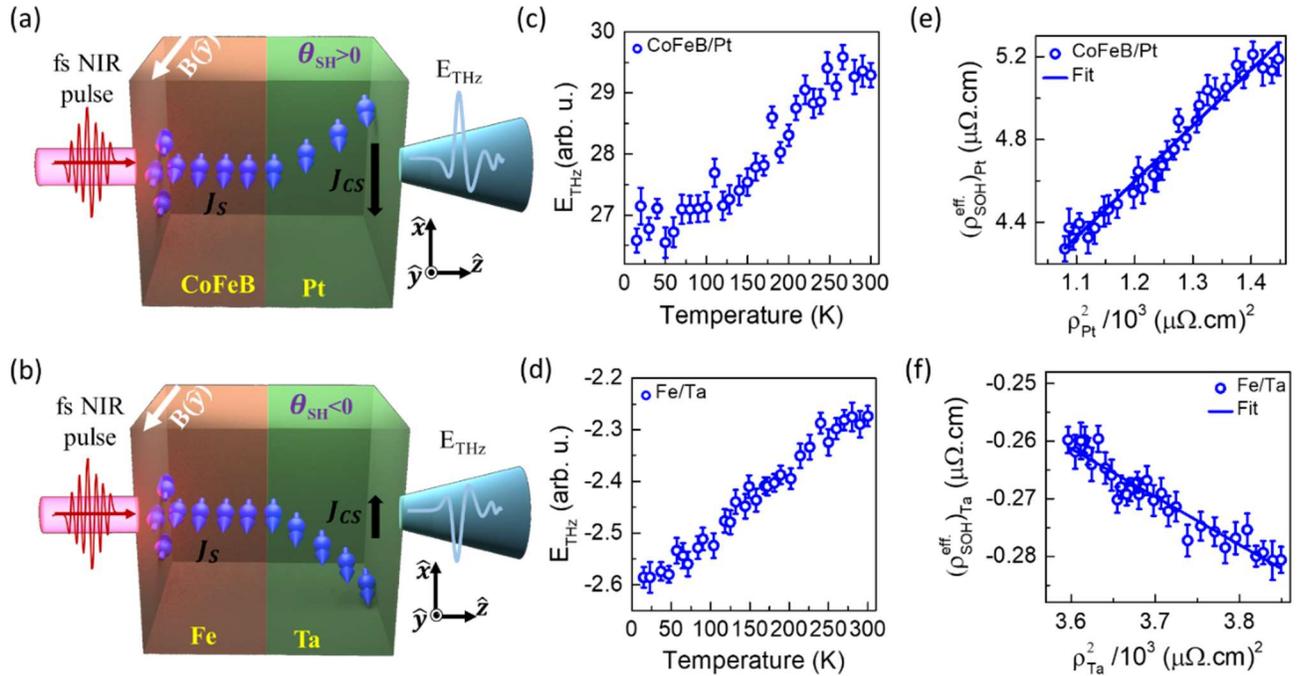

**Fig. 3. Temperature-dependence of THz signal and effective Hall resistivity for CoFeB/Pt and Fe/Ta bilayers (Pt, Ta = heavy metal).** Schematic illustration of the ultrafast optical excitation, spin magnetic moment transport and its conversion to transient charge current through ISHE to emit a THz pulse from (a) CoFeB/Pt and (b) Fe/Ta. The spin Hall angle ($\theta_{SH}$) and the direction of the external magnetic field are indicated. Temperature-dependent variation of the THz signal peak-to-peak amplitude for (c) CoFeB/Pt, and (d) Fe/Ta. Effective spin-orbital Hall resistivity, $\rho_{SOH}^{eff}$ as a function of the squared longitudinal resistivity, $\rho_{NM}^2$ for (e) Pt and (f) Ta. The continuous curves in (e) and (f) are linear fits to the data using Eq. (4).

## 2.3 CoFeB/W/Ta: A heavy metal insertion layer enhances orbital transport and hence the THz generation efficiency

In the above two sections, we have established role of IOHE in NiFe/Nb and ISHE in CoFeB/Pt and Fe/Ta, as the principal reason for the generation of THz pulses from them. Due to comparable or even larger value of $\sigma_{OH}$ than $\sigma_{SH}$ in some heavy metals, the existence of significant OHE or IOHE is also expected in them. But to harness the effect for its direct observation, one needs to choose their appropriate combinations with the FM layers. To launch an orbital current into the heavy metal layer of a FM/NM structure, use[54] of Ni or NiFe, or similar other FM materials, would be the proper choice; otherwise, despite of large $\sigma_{OH}$ in some heavy metals, for example, nearly an order higher $\sigma_{OH}$ than $\sigma_{SH}$ in Ta, it is very difficult to observe orbital transport in either Fe/Ta or CoFeB/Pt, the two cases described in the previous section. In this section, we show that by adding/interfacing a heavy metal W-insertion layer in technologically relevant CoFeB/Ta heterostructure, the orbital transport gets pronounced. For this study, CoFeB(2)/W(2)/Ta(2) and CoFeB(2)/W(1)/Ta(2), where, the integers inside small parentheses represent layer thickness in nm, were fabricated. The thickness and good interface quality of the W-insertion layer is confirmed by analysing the elemental stack using secondary ion mass spectroscopy (SIMS) technique, as shown in Supplementary Section S2. The thickness constraint on the W-insertion layer is motivated by the previous studies.[71,72] We may emphasize that by adding the W-insertion layer, the magnetic properties of the trilayers are nearly unchanged from the bilayer counterparts, as confirmed from both the MH and MT-measurements (see Sections S3 and S4 of the Supplementary Information).

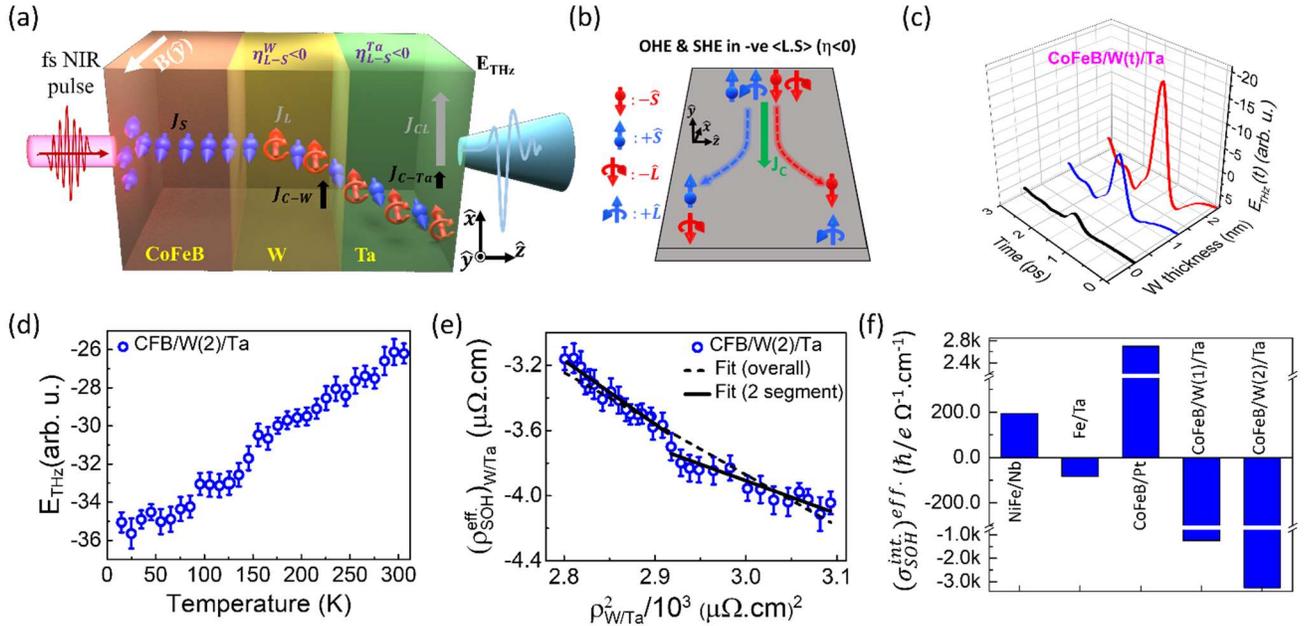

**Fig. 4. Enhanced spin-to-orbital current conversion by W-insertion layer in CoFeB/W/Ta.** (a) Schematic to illustrate spin to orbital current conversion due to large negative spin-orbit correlation factor, $\eta_{L-S}^W (<0)$ in the W-insertion layer of the CoFeB/W/Ta heterostructure which is ultrafast optically excited from the substrate side. The transient charge currents due to spin-charge conversion in W and Ta layers are labelled as $J_{C-W}$ and $J_{C-Ta}$, respectively, while, the charge current in Ta layer due to orbital-charge conversion is labelled as $J_{CL}$. (b) Schematic illustration of transport of spin and orbital moments in a material having negative spin-orbit correlation ($\eta_{L-S} < 0$) such as W and Ta. (c) Time-domain traces of the THz radiation emitted by CoFeB/W(t)/Ta heterostructure with varying thickness of the W-insertion layer, t = 0, 1, 2 nm. (d) Peak-to-peak THz signal amplitude variation with respect to the sample temperature for CoFeB/W(2)/Ta. (e) Effective spin-orbital Hall resistivity, $\rho_{SOH}^{eff.}$ as a function of the squared longitudinal resistivity, $\rho_{W/Ta}^2$. The solid and dashed lines represent linear fits as discussed in the text. (f) Extracted values of effective intrinsic spin-orbital Hall conductivity, $(\sigma_{SOH}^{int})^{eff}$ for different bi- and trilayer heterostructures used in the current study.

Figure 4(a) schematically shows the ultrafast optically excited spin current in the CoFeB layer and subsequent injection into the W-insertion layer. Heavy metal W supports efficient spin-orbit conversion.[54] Thus, within the W-insertion layer, in addition to a fractional spin-charge conversion via ISHE, an orbital current is also produced from the incoming spin current. Due to the high and negative value of the spin-orbit correlation factor $\eta_{L-S}^W$ of W, as explained in Fig. 4(b), the spin and orbital moments are oppositely polarized during their propagation in the W-insertion layer. Subsequently, both the spin and orbital currents are injected from the W-insertion layer into the adjacent heavy metal Ta layer. Ta is known to possess[8,14] nearly an order higher and positive value of $\sigma_{OH}$ than the $\sigma_{SH}$ (see Table 1), and the corresponding spin-Hall angle is also much smaller in the α-phase Ta. Therefore, maximal charge current conversion from the orbital current via IOHE and comparatively negligible from the spin current via ISHE takes place in the Ta layer. The net charge current produced in the CoFeB/W/Ta trilayer that is responsible for THz emission from it can be expressed by a relation like Eq. (1) as follows,

$$J_C = J_{C-IS} + J_{C-I} = (\theta_{SH}^{Ta}.J_s + \theta_{SH}^{W}.J_s) + \theta_{OH}^{Ta}.\eta_{L-S}^{W}.J_S \qquad (6)$$

Here, the signs of $\theta_{SH}^{Ta}$, $\theta_{SH}^{W}$, $\eta_{L-S}^{W}$ are negative,[54,73] while it is positive for $\theta_{OH}^{Ta}$. Therefore, the transient charge currents add up constructively to generate enhanced coherent THz radiation.

Figure 4(c) presents time-domain traces of the THz radiation emitted from CoFeB(2)/W(t)/Ta(2) trilayer samples of varying W-insertion layer thickness from t = 0 to 2 nm. While the THz amplitude for t = 2 nm is nearly 10 times stronger than that from t = 0 nm, i.e., the CoFeB(2)/Ta(2) bilayer, the same is nearly 4 times higher than from the CoFeB(2)/W(2) bilayer (see Supplementary Section S14). The monotonic increase in the THz signal amplitude with the thickness of the W-insertion layer in Fig. 4(c) is in contradiction to a previous study[39] where, irrespective of the type of insertion layer material, a decrease in the THz emission efficiency with the increasing insertion layer thickness is reported. There, the results were interpreted in the context of increased spin memory loss in the insertion layer. At the same time, in few other reports in the recent literature,[71,72,74,75] significant enhancement in the spin current flow due to an insertion layer was attributed to the atomically thin nature of the insertion layer. We argue that such inconsistency in the literature is arising because of the fact that conventionally, the experimental outcomes have been interpreted solely in terms of the spin current, neglecting the effect of orbital contribution altogether, though the latter can be gigantic even in the light metals with weak SOC.[13,14] The orbital current diffusion length can be several times larger than the spin counterpart in some heavy metals.[7,25,48,53,59] The orbital current due to spin-orbit conversion within the heavy metal W-insertion layer is given as[48] $J_L \propto \int_0^d J_S \eta_{L-S}^{W} dt$, according to which, higher thickness and stronger $\eta$, both contribute to enhance the orbital current (see Supplementary Section S16). Here, the orbital current diffusion length[54] in W is much larger than the thicknesses of the insertion layer used currently. These reasons clearly justify our observation of W-insertion layer thickness dependent enhancement in the THz emission via IOHE in CoFeB/W/Ta. The nearly one order increase in the THz generation efficiency of CoFeB/W/Ta relative to CoFeB/Ta, as shown in Fig. 4(c) can be clearly attributed to the strong orbital current injection from the W layer into the Ta layer and subsequently large orbital-charge conversion via IOHE in it.

To strengthen our point that indeed the orbital current within the W-insertion layer contributes to enhance the THz pulse emission from CoFeB/W/Ta as compared to CoFeB/Ta bilayer counterpart, we now present temperature-dependent results on them. The governing role of IOHE for THz emission from CoFeB/W/Ta, due to the orbital current within the W-insertion layer that increases in proportion with its thickness, is brought out clearly. In Fig. 4(d), we have plotted the variation in the THz signal amplitude as a function of CoFeB/W(2)/Ta sample temperature. The corresponding longitudinal resistivities are also recorded at each sample temperature and presented in Supplementary Section S7. The resistivity information, together with the THz amplitude data in Fig. 4(d), are used to obtain the results presented in Fig. 4(e), where the extracted effective Hall resistivity ($\rho_{SOH}^{eff}$) with respect to the squared longitudinal electrical resistivity ($\rho_{W/Ta}^2$) is plotted at each temperature following the same procedure as discussed earlier in the paper. Additional temperature-dependent results obtained on CoFeB/W(1)/Ta sample for the different thickness of the W-insertion layer are included in the Supplementary Section S9, where more or less same temperature trends have been obtained. We have fitted the experimental data for CoFeB/W(2)/Ta in Fig. 4(e) using Eq. (4) to obtain the effective intrinsic spin-orbital Hall conductivity $(\sigma_{SOH}^{int})^{eff}$ from the slope of the linear fit. The behavior of the data suggests that a two-segment linear curve fitting would be more appropriate here. Though, the degree of linearity is enhanced with two-segment linear curve fitting as compared to the single linear curve fitting, yet both approaches yield nearly the same value of the slope and there is no clear distinction in terms of low and high-resistivity regions as was the case with NiFe/Nb, discussed earlier. Therefore, similar to the cases of AHE[66-69] and SHE[63], the OHE here is expected to be dominated by the intrinsic contribution in the entire temperature range of the Hall resistivity. From the linear fitting, we obtain $(\sigma_{SOH}^{int})^{eff} = (\sigma_{SH}^{int} + \eta_{L-S}.\sigma_{OH}^{int}) \sim -3256\ (\hbar/e)\ \Omega^{-1}cm^{-1}$. Given that the values of the intrinsic spin Hall conductivities[15] of both α-phase W[76] and Ta[73] are $(\sigma_{SH}^{int})^{Ta} \sim -700\ (\hbar/e)\ \Omega^{-1}cm^{-1}$, and $(\sigma_{SH}^{int})^{W} \sim -103\ (\hbar/e)\ \Omega^{-1}cm^{-1}$, therefore, the large and negative value of $(\sigma_{SOH}^{int})^{eff} \sim -3256\ (\hbar/e)\ \Omega^{-1}cm^{-1}$ clearly suggests that the term, $\eta_{L-S}.\sigma_{OH}^{int}$, must be of a relatively much larger negative value. From the available literature, we find that indeed the intrinsic orbital Hall conductivity in Ta dominates[8,18] over the intrinsic spin Hall conductivity ($\sigma_{OH}^{int} \gg \sigma_{SH}^{int}$), thus, signifying the pronounced orbital current injection from W-insertion layer and its conversion to transient charge current in Ta layer of CoFeB/W/Ta heterostructure. Nearly half the value of $(\sigma_{SOH}^{int})^{eff} \sim -1250\ (\hbar/e)\ \Omega^{-1}cm^{-1}$ is obtained by analyzing the data for CoFeB/W(1)/Ta (see Supplementary Information), where the magnitude of the THz signal gets reduced in proportion with the thickness of the W-insertion layer. Therefore, the temperature dependent analysis of CoFeB/W/Ta samples strengthens our argument that the enhanced THz signal from CoFeB/W/Ta as compared to the CoFeB/Ta bilayer is due to the strong spin-to-orbit conversion within the W-insertion layer and subsequently efficient orbit-charge conversion via IOHE in the Ta layer of the CoFeB/W/Ta heterostructure.

Figure 4(f) summarizes our results on the effective intrinsic spin-orbital Hall conductivity of different heterolayer systems that are experimentally determined in a non-contact and non-invasive manner by time-domain THz emission spectroscopy. To have a ready comparison of the values for different materials and heterostructures, available from different theoretical and experimental studies in the literature, Table 1 lists them together with the values of the spin-orbit correlation factor, wherever applicable.

**TABLE. 1.** Comparison of the spin ($\sigma_{SH}$), orbital ($\sigma_{OH}$), and effective Hall $(\sigma_{SOH}^{int})^{eff}$ conductivities in different materials and heterostructures. Values of the spin-orbit correlation factor, $\eta_{L-S}$, wherever available, are also listed.

| Materials and heterostructures | $\sigma_{SH}$ ($\Omega^{-1}.cm^{-1}$) | $\sigma_{OH}$ ($\Omega^{-1}.cm^{-1}$) | $(\sigma_{SOH}^{int})^{eff}$ ($\Omega^{-1}.cm^{-1}$) | $\eta_{L-S}$ |
|---|---|---|---|---|
| Pt | +2152[8] +815[14] +1739[18] | +2919[8] +1918[14] +4330[18] | | +ve |
| CoFeB/Pt | -- | -- | +2208[8] +2140[8] +2704 (this work) | -- |
| Ta | -274[8] -286[14] -7[18] | +6803[8] +3820[14] +3950[18] | | -ve |
| Co/Ta | -- | -- | -98[8] | -- |
| FeB/Ta | -- | -- | -359[8] | -- |
| Fe/Ta | -- | -- | -211[8] -83 (this work) | -- |
| W | -324[14] -83[18] | +3293[14] +4490[18] | | -ve |
| CoFeB/W(1)/Ta | -- | -- | -1250 (this work) | -- |
| CoFeB/W(2)/Ta | -- | -- | -3256 (this work) | -- |
| Nb | -117[14] -48[18] | +3641[14] +5930[18] | | -ve |
| Ni/Nb | -- | -- | +293[59] | -- |
| Co/Nb | -- | -- | +149[60] | -- |
| NiFe/Nb | -- | -- | +195 (this work) | -- |

## 3. CONCLUSION

In summary, we have experimentally demonstrated the ultrafast optically induced orbital current and its conversion to transient charge current via IOHE by wide-range temperature-dependent THz emission measurements on multiple FM/NM heterostructures. To show the role of the spin and orbital currents exclusively, we have chosen the material combinations in the layered heterostructures such that they comprise either a heavy or a light metal layer. THz pulses emitted from them have been measured as a function of the sample temperature to disentangle the contributions from the spin and the orbital transports. Majorly the orbital-charge conversion via IOHE in NiFe/Nb and the spin-charge conversion via ISHE in CoFeB/Pt, CoFeB/W, CoFeB/Ta and Fe/Ta, are manifested from the extracted values of effective intrinsic Hall conductivities. Further analysis of NiFe/Nb system reveals signature of different resistivity regimes dominated by either intrinsic or extrinsic contributions to OHE, which has not been seen before in any orbital Hall systems. We also find that an insertion layer of heavy metal W in the CoFeB/W/Ta heterostructure, provides a pathway to constitute an ultrafast orbital current within it and subsequently its conversion to transient charge current in the Ta layer via IOHE that significantly enhances the THz generation efficiency as compared to the CoFeB/Ta counterpart. These findings will be proven to be highly useful in efforts towards realizing ultrafast orbitronic devices as well as adding new knowledge of the underlying physics.

**METHODS**

Heterostructures comprising of ferromagnetic $Co_{20}Fe_{60}B_{20}$ (CoFeB) and $Ni_{90}Fe_{10}$ (NiFe) layers, and nonmagnetic Pt (platinum), Ta (tantalum), W (tungsten), Nb (niobium) material layers were created by using ultra high vacuum radio frequency magnetron sputtering technique. Bilayer systems, Sub./CoFeB(2)/Pt(3), Sub./CoFeB(2)/Ta(2), Sub./Fe(2)/Ta(3) and Sub./NiFe(5)/Nb(10), and trilayer systems, Sub./CoFeB(2)/W(2)/Ta(2) and Sub./CoFeB(2)/W(1)/Ta(2), were deposited layer by layer on 1 mm thick quartz substrates (Sub.). The information related to the film thickness, roughness and phase is obtained by using various structural and topographical measurements techniques, such as X-ray diffraction (XRD), X-ray reflectivity (XRR), and atomic force microscopy (AFM), and can be found in the Supplementary Section S1. SIMS depth profile experiments reconfirmed the elemental stack and the quality of interfaces in the heterostructures (see Supplementary Section S2). The magnetic measurements shown in Supplementary Sections S3 and S4, were performed in a magnetic properties measurements system (MPMS3, Quantum Design). We have employed a closed-cycle helium optical cryostat system (SHI-4-2-XG, Janis) operating in the temperature range of 10–450 K for all the temperature-dependent electrical transport and time-domain THz experiments. Complete details of the temperature-dependent THz setup[61] are provided in the Supplementary Section S5. A regenerative femtosecond amplifier (Astrella, Coherent Inc.) providing laser pulses of ~50 fs pulse duration at 1 kHz repetition rate and centred at 800 nm wavelength, were used for the THz generation and detection. The collimated optical excitation (pump) beam diameter on the sample was kept at ~3 mm. All the samples were excited from the substrate side, unless specified. The emitted THz pulses were collected from behind the sample by a set of two $90^0$ off-axis gold-coated parabolic mirrors of focal length of 15 cm. THz pulses were detected by electro-optic sampling scheme in a (110)-oriented ZnTe crystal of thickness 500 micron by using a combination of a quarter wave plate, a Wollaston prism, a balanced photodiode and a lock-in amplifier.[61] The THz setup was under the normal conditions of the room temperature and humidity.

## DATA AVAILABILITY

Source data are provided with this paper.

## ACKNOWLEDGMENTS

SK acknowledges the Science and Engineering Research Board (SERB), Department of Science and Technology, Government of India, for financial support through project no. CRG/2020/000892, Joint Advanced Technology Center, IIT Delhi, is also acknowledged for support through EMDTERA#5 project. The authors acknowledge CRF, IIT Delhi for SIMS facilities. One of the authors (Sandeep Kumar) acknowledges the University Grants Commission, Government of India, for Senior Research Fellowship.

## AUTHOR CONTRIBUTIONS

Sunil Kumar supervised the work. Sandeep Kumar performed the experiments. Sunil Kumar and Sandeep Kumar analyzed the data and wrote the manuscript.

## COMPETING INTERESTS

The authors declare no competing interests.

## ADDITIONAL INFORMATION

Supporting Information is available. (Attached below)

## REFERENCES

1  Hirohata, A. *et al.* Review on spintronics: Principles and device applications. *Journal of Magnetism and Magnetic Materials* **509**, 166711, (2020).
2  Dieny, B. *et al.* Opportunities and challenges for spintronics in the microelectronics industry. *Nature Electronics* **3**, 446-459, (2020).
3  Sinova, J., Valenzuela, S. O., Wunderlich, J., Back, C. H. & Jungwirth, T. Spin Hall effects. *Reviews of Modern Physics* **87**, 1213-1260, (2015).
4  Hirsch, J. E. Spin Hall Effect. *Physical Review Letters* **83**, 1834-1837, (1999).
5  Manchon, A., Koo, H. C., Nitta, J., Frolov, S. M. & Duine, R. A. New perspectives for Rashba spin–orbit coupling. *Nature Materials* **14**, 871-882, (2015).
6  Edelstein, V. M. Spin polarization of conduction electrons induced by electric current in two-dimensional asymmetric electron systems. *Solid State Communications* **73**, 233-235, (1990).
7  Lee, S. *et al.* Efficient conversion of orbital Hall current to spin current for spin-orbit torque switching. *Communications Physics* **4**, 234, (2021).
8  Lee, D. *et al.* Orbital torque in magnetic bilayers. *Nature Communications* **12**, 6710, (2021).
9  Go, D., Jo, D., Lee, H.-W., Kläui, M. & Mokrousov, Y. Orbitronics: Orbital currents in solids. *Europhysics Letters* **135**, 37001, (2021).
10 Zhang, S. & Yang, Z. Intrinsic Spin and Orbital Angular Momentum Hall Effect. *Physical Review Letters* **94**, 066602, (2005).

11  Bernevig, B. A., Hughes, T. L. & Zhang, S.-C. Orbitronics: The Intrinsic Orbital Current in p-Doped Silicon. *Physical Review Letters* **95**, 066601, (2005).
12  Kittel, C. *Introduction to solid state physics*. 8 edn, (John Wiley & Sons, 2004).
13  Jo, D., Go, D. & Lee, H.-W. Gigantic intrinsic orbital Hall effects in weakly spin-orbit coupled metals. *Physical Review B* **98**, 214405, (2018).
14  Kontani, H., Tanaka, T., Hirashima, D. S., Yamada, K. & Inoue, J. Giant Orbital Hall Effect in Transition Metals: Origin of Large Spin and Anomalous Hall Effects. *Physical Review Letters* **102**, 016601, (2009).
15  Tanaka, T. *et al.* Intrinsic spin Hall effect and orbital Hall effect in 4d and 5d transition metals. *Physical Review B* **77**, 165117, (2008).
16  Choi, Y.-G. *et al.* Observation of the orbital Hall effect in a light metal Ti. *Nature* **619**, 52-56, (2023).
17  Sala, G. & Gambardella, P. Giant orbital Hall effect and orbital-to-spin conversion in 3d, 5d, and 4f metallic heterostructures. *Physical Review Research* **4**, 033037, (2022).
18  Salemi, L. & Oppeneer, P. M. First-principles theory of intrinsic spin and orbital Hall and Nernst effects in metallic monoatomic crystals. *Physical Review Materials* **6**, 095001, (2022).
19  Go, D. & Lee, H.-W. Orbital torque: Torque generation by orbital current injection. *Physical Review Research* **2**, 013177, (2020).
20  Go, D., Jo, D., Kim, C. & Lee, H.-W. Intrinsic Spin and Orbital Hall Effects from Orbital Texture. *Physical Review Letters* **121**, 086602, (2018).
21  Sahu, P., Bhowal, S. & Satpathy, S. Effect of the inversion symmetry breaking on the orbital Hall effect: A model study. *Physical Review B* **103**, 085113, (2021).
22  Bhowal, S. & Satpathy, S. Intrinsic orbital moment and prediction of a large orbital Hall effect in two-dimensional transition metal dichalcogenides. *Physical Review B* **101**, 121112, (2020).
23  Cysne, T. P. *et al.* Disentangling Orbital and Valley Hall Effects in Bilayers of Transition Metal Dichalcogenides. *Physical Review Letters* **126**, 056601, (2021).
24  Baek, I. & Lee, H.-W. Negative intrinsic orbital Hall effect in group XIV materials. *Physical Review B* **104**, 245204, (2021).
25  Santos, E. *et al.* Inverse Orbital Torque via Spin-Orbital Intertwined States. *Physical Review Applied* **19**, 014069, (2023).
26  Go, D. *et al.* Theory of current-induced angular momentum transfer dynamics in spin-orbit coupled systems. *Physical Review Research* **2**, 033401, (2020).
27  Kimura, T., Otani, Y., Sato, T., Takahashi, S. & Maekawa, S. Room-Temperature Reversible Spin Hall Effect. *Physical Review Letters* **98**, 156601, (2007).
28  Papaioannou, E. T. & Beigang, R. THz spintronic emitters: a review on achievements and future challenges. *Nanophotonics* **10**, 1243-1257, (2021).
29  Kumar, S. *et al.* Optical damage limit of efficient spintronic THz emitters. *iScience* **24**, 103152, (2021).
30  Torosyan, G., Keller, S., Scheuer, L., Beigang, R. & Papaioannou, E. T. Optimized Spintronic Terahertz Emitters Based on Epitaxial Grown Fe/Pt Layer Structures. *Scientific Reports* **8**, 1311, 2018).
31  Yang, D. *et al.* Powerful and Tunable THz Emitters Based on the Fe/Pt Magnetic Heterostructure. *Advanced Optical Materials* **4**, 1944-1949, (2016).
32  Seifert, T. *et al.* Efficient metallic spintronic emitters of ultrabroadband terahertz radiation. *Nature Photonics* **10**, 483, (2016).
33  Kampfrath, T. *et al.* Terahertz spin current pulses controlled by magnetic heterostructures. *Nature Nanotechnology* **8**, 256, (2013).
34  Kumar, S., Nivedan, A., Singh, A. & Kumar, S. THz pulses from optically excited Fe-, Pt- and Ta-based spintronic heterostructures. *Pramana* **95**, 75, (2021).
35  Zhou, C. *et al.* Broadband Terahertz Generation via the Interface Inverse Rashba-Edelstein Effect. *Physical Review Letters* **121**, 086801, (2018).
36  Jungfleisch, M. B. *et al.* Control of Terahertz Emission by Ultrafast Spin-Charge Current Conversion at Rashba Interfaces. *Physical Review Letters* **120**, 207207, (2018).
37  Qiu, H. *et al.* Ultrafast spin current generated from an antiferromagnet. *Nature Physics* **17**, 388-394, (2021).
38  Seifert, T. *et al.* Ultrabroadband single-cycle terahertz pulses with peak fields of 300 kV cm−1 from a metallic spintronic emitter. *Applied Physics Letters* **110**, 252402, (2017).
39  Hawecker, J. *et al.* Spin Injection Efficiency at Metallic Interfaces Probed by THz Emission Spectroscopy. *Advanced Optical Materials* **9**, 2100412, (2021).
40  Gueckstock, O. *et al.* Terahertz Spin-to-Charge Conversion by Interfacial Skew Scattering in Metallic Bilayers. *Advanced Materials* **33**, 2006281, (2021).
41  Cheng, L., Li, Z., Zhao, D. & Chia, E. E. M. Studying spin–charge conversion using terahertz pulses. *APL Materials* **9**, 070902, (2021).
42  Agarwal, P. *et al.* Ultrafast Photo-Thermal Switching of Terahertz Spin Currents. *Advanced Functional Materials* **31**, 2010453, (2021).
43  Dang, T. H. *et al.* Ultrafast spin-currents and charge conversion at 3d-5d interfaces probed by time-domain terahertz spectroscopy. *Applied Physics Reviews* **7**, 041409, (2020).
44  Seifert, T. S. *et al.* Terahertz spectroscopy for all-optical spintronic characterization of the spin-Hall-effect metals Pt, W and Cu80Ir20. *Journal of Physics D: Applied Physics* **51**, 364003, (2018).
45  Gorchon, J., Mangin, S., Hehn, M. & Malinowski, G. Is terahertz emission a good probe of the spin current attenuation length? *Applied Physics Letters* **121**, 012402, (2022).
46  Kumar, S. & Kumar, S. Ultrafast light-induced THz switching in exchange-biased Fe/Pt spintronic heterostructure. *Applied Physics Letters* **120**, 202403, (2022).
47  Bull, C. *et al.* Spintronic terahertz emitters: Status and prospects from a materials perspective. *APL Materials* **9**, 090701, (2021).
48  Xu, Y. *et al.* Inverse Orbital Hall Effect Discovered from Light-Induced Terahertz Emission. arXiv:2208.01866 (2022).
49  Wang, P. *et al.* Inverse orbital Hall effect and orbitronic terahertz emission observed in the materials with weak spin-orbit coupling. *npj Quantum Materials* **8**, 28, (2023).

50  Shi, Z. *et al.* Effect of band filling on anomalous Hall conductivity and magneto-crystalline anisotropy in NiFe epitaxial thin films. *AIP Advances* **6**, 015101, (2016).
51  Stamm, C., Pontius, N., Kachel, T., Wietstruk, M. & Dürr, H. A. Femtosecond x-ray absorption spectroscopy of spin and orbital angular momentum in photoexcited Ni films during ultrafast demagnetization. *Physical Review B* **81**, 104425, (2010).
52  Rouzegar, R. *et al.* Laser-induced terahertz spin transport in magnetic nanostructures arises from the same force as ultrafast demagnetization. *Physical Review B* **106**, 144427, (2022).
53  Seifert, T. S. *et al.* Time-domain observation of ballistic orbital-angular-momentum currents with giant relaxation length in tungsten. *Nature Nanotechnology* (2023).
54  Hayashi, H. *et al.* Observation of long-range orbital transport and giant orbital torque. *Communications Physics* **6**, 32, (2023).
55  Go, D. *et al.* Long-Range Orbital Torque by Momentum-Space Hotspots. *Physical Review Letters* **130**, 246701, (2023).
56  Boeglin, C. *et al.* Distinguishing the ultrafast dynamics of spin and orbital moments in solids. *Nature* **465**, 458-461, (2010).
57  Stamm, C. *et al.* Femtosecond modification of electron localization and transfer of angular momentum in nickel. *Nature Materials* **6**, 740-743, (2007).
58  Hennecke, M. *et al.* Angular Momentum Flow During Ultrafast Demagnetization of a Ferrimagnet. *Physical Review Letters* **122**, 157202, (2019).
59  Bose, A. *et al.* Detection of long-range orbital-Hall torques. *Physical Review B* **107**, 134423, (2023).
60  Liu, F., Liang, B., Xu, J., Jia, C. & Jiang, C. Giant efficiency of long-range orbital torque in Co/Nb bilayers. *Physical Review B* **107**, 054404, (2023).
61  Kumar, S. & Kumar, S. Large interfacial contribution to ultrafast THz emission by inverse spin Hall effect in CoFeB/Ta heterostructure. *iScience* **25**, 104718, (2022).
62  Sagasta, E. *et al.* Unveiling the mechanisms of the spin Hall effect in Ta. *Physical Review B* **98**, 060410, (2018).
63  Sagasta, E. *et al.* Tuning the spin Hall effect of Pt from the moderately dirty to the superclean regime. *Physical Review B* **94**, 060412, (2016).
64  Matthiesen, M. *et al.* Temperature dependent inverse spin Hall effect in Co/Pt spintronic emitters. *Applied Physics Letters* **116**, 212405, (2020).
65  Fert, A. & Levy, P. M. Spin Hall Effect Induced by Resonant Scattering on Impurities in Metals. *Physical Review Letters* **106**, 157208, (2011).
66  Sangiao, S. *et al.* Anomalous Hall effect in Fe (001) epitaxial thin films over a wide range in conductivity. *Physical Review B* **79**, 014431, (2009).
67  Onoda, S., Sugimoto, N. & Nagaosa, N. Quantum transport theory of anomalous electric, thermoelectric, and thermal Hall effects in ferromagnets. *Physical Review B* **77**, 165103, (2008).
68  Onoda, S., Sugimoto, N. & Nagaosa, N. Intrinsic Versus Extrinsic Anomalous Hall Effect in Ferromagnets. *Physical Review Letters* **97**, 126602, (2006).
69  Tian, Y., Ye, L. & Jin, X. Proper Scaling of the Anomalous Hall Effect. *Physical Review Letters* **103**, 087206, (2009).
70  Suess, D., Fidler, J., Zimanyi, G., Schrefl, T. & Visscher, P. Thermal stability of graded exchange spring media under the influence of external fields. *Applied Physics Letters* **92**, 173111, (2008).
71  Li, Y. *et al.* Enhancing the Spin–Orbit Torque Efficiency by the Insertion of a Sub-nanometer β-W Layer. *ACS Nano* **16**, 11852-11861, (2022).
72  Lu, Q. *et al.* Enhancement of the Spin-Mixing Conductance in CoFeB/W Bilayers by Interface Engineering. *Physical Review Applied* **12**, 064035, (2019).
73  Kumar, A., Bansal, R., Chaudhary, S. & Muduli, P. K. Large spin current generation by the spin Hall effect in mixed crystalline phase Ta thin films. *Physical Review B* **98**, 104403, (2018).
74  Wahada, M. A. *et al.* Atomic Scale Control of Spin Current Transmission at Interfaces. *Nano Letters* **22**, 3539-3544, (2022).
75  Hait, S. *et al.* Spin Pumping through Different Spin–Orbit Coupling Interfaces in β-W/Interlayer/Co2FeAl Heterostructures. *ACS Applied Materials & Interfaces* **14**, 37182-37191, (2022).
76  McHugh, O. L. W., Goh, W. F., Gradhand, M. & Stewart, D. A. Impact of impurities on the spin Hall conductivity in β-W. *Physical Review Materials* **4**, 094404, (2020).

# Supplementary Information

# Ultrafast THz probing of nonlocal orbital current in transverse multilayer metallic heterostructures

Sandeep Kumar and Sunil Kumar*

*Femtosecond Spectroscopy and Nonlinear Photonics Laboratory,
Department of Physics, Indian Institute of Technology Delhi, New Delhi 110016, India
*Email: kumarsunil@physics.iitd.ac.in*

## S1. Structural characterization of the samples

Figure S1 presents the results from X-ray diffraction (XRD), X-ray reflectivity (XRR), and atomic force microscopy (AFM) for determining the crystalline phase, thickness, roughness, and other characteristics of the thin film samples. The X-ray measurements were carried out using PANalytical X'Pert diffractometer with a Cu-K$_\alpha$ source. The XRD pattern of NiFe/Nb bilayer is shown in Fig. S1(a), where strong diffraction peaks at $38.6^0$, $48.4^0$, $55.4^0$, and $69.3^0$, corresponding to the crystalline planes of NiFe and Nb are marked. These results show that the Nb film is grown in the body centered cubic form while NiFe is grown in the face centered cubic form. To achieve growth of the W and Ta films in their low resistive α-phase, optimized growth rates of 0.05 nm/sec(Ta) and 0.03 nm/sec(W) were used while deposition power was kept constant[1,2] at 50 W. A balanced growth rate along with the optimized deposition power is necessary to obtain the desired phase formation[3,4]. Although all the samples were grown using the respective optimized growth rates for all the materials, the film/layer thickness and roughness were further confirmed by performing XRR. Figure S1(b) shows the XRR profile of NiFe/Nb bilayer sample. The experimental data in green open circles is fitted using GenX software to obtain information about film thickness and roughness. The AFM image as showin in Fig. S1(c) further confirms the surface morphology and roughness of the same sample. Root mean squared roughness extracted from the AFM result is found to be consistent with the one obtained from the XRR measurements. A small surface roughness incontrovertibly indicates very smooth sample quality. Other samples, such as CoFeB/W(1)/Ta (Fig. S1(d)), CoFeB/W(2)/Ta (Fig. S1(e)), were also examined in the similar fashion by utilizing XRR for their thickness and roughnesses in addition to the complementary AFM (Fig. S1(f)). Again, a lower value of surface and interfacial roughness confirms the very good quality of our prepared thin film samples.

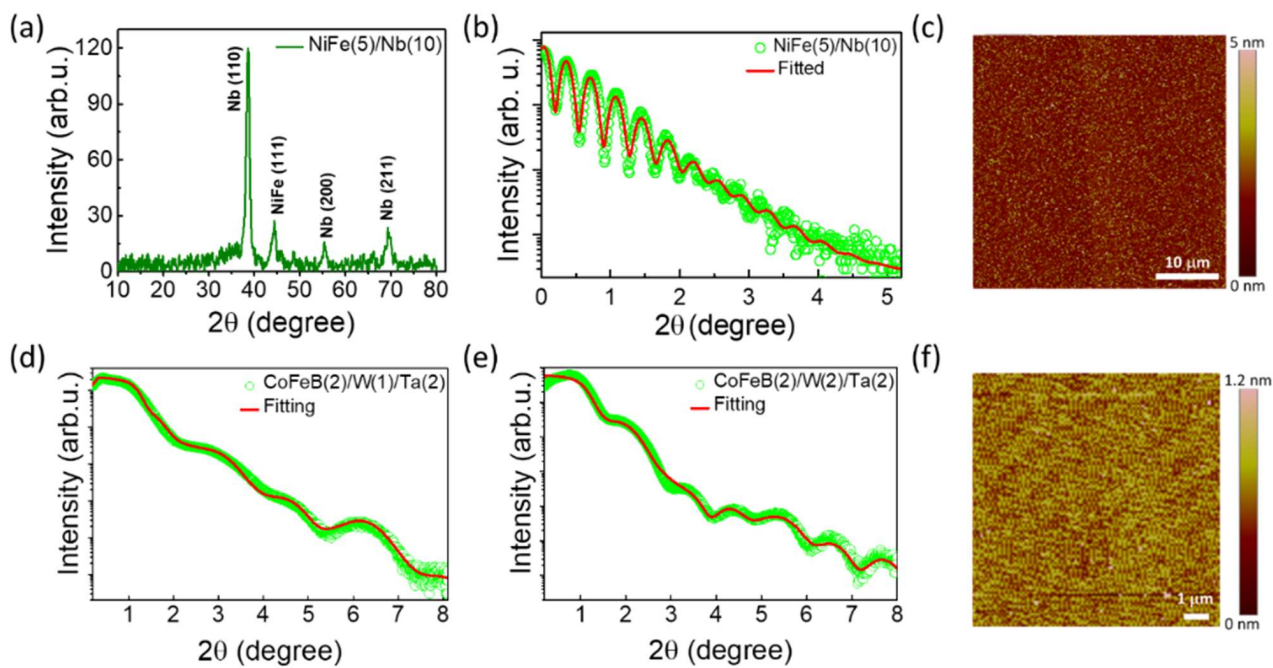

**Figure S1.** (a) XRD, (b) XRR, (c) AFM results on the NiFe/Nb bilayer sample. The crystallographic planes corresponding to various XRD peaks have been marked. XRR spectra of (d) CoFeB/W(1)/Ta and (e) CoFeB/W(2)/Ta trilayer samples. Numbers inside the small parentheses indicate film thickness in nm. (f) AMF image of CoFeB/W(2)/Ta sample. The solid red curves in (b,d,e) are fits to the experimental data for obtaining film thickness and surface roughness.

## S2. Secondary ion mass spectroscopy (SIMS) measurements

To reconfirm the elemental composition over the thickness as well as the interface quality, we have performed secondary ion mass spectroscopy (SIMS) on a few selected samples, including CoFeB(2)/Ta(2), CoFeB(2)/W(2)/Ta(2) and NiFe(5)/Nb(10). The number inside the small parenthesis represents the film thickness in nm. A $O_2^+$ ion source of energy 0.5 keV was used for the depth profile measurements. The results are shown in Fig. S2, where the yield, referred as normalized intensity, is plotted with respect to the sample thickness from bottom to the top order of the stack. It is evident from the depth profiles that the elemental compositions of a given layer covered a range of thicknesses, and this thickness is consistent with that determined from the structural measurements made in section S1. For example, in the case of NiFe(5)/Nb(10) (Fig. S2(a)), the Ni and Fe elements span a thickness of 5 nm, while the Nb starts after this and ends at a thickness of ~15 nm. By comparing the results in Figs. S2(b) and S2(c), thickness of 2 nm is confirmed for the W-insertion layer.

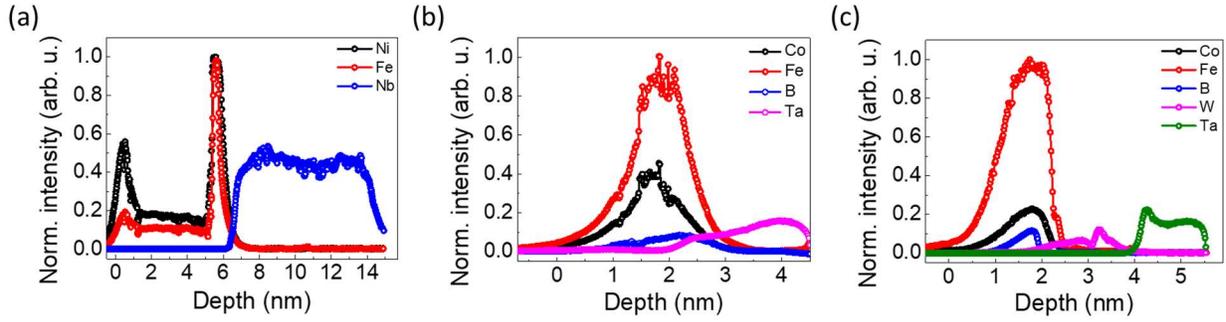

**Figure S2.** Secondary ion mass spectroscopy depth profiles of various elements in the thin film heterostructures of (a) NiFe(5)/Nb(10), (b) CoFeB(2)/Ta(2), and (c) CoFeB(2)/W(2)/Ta(2). The numbers inside small parentheses represent the corresponding film thickness in the nm.

## S3. M-H measurements

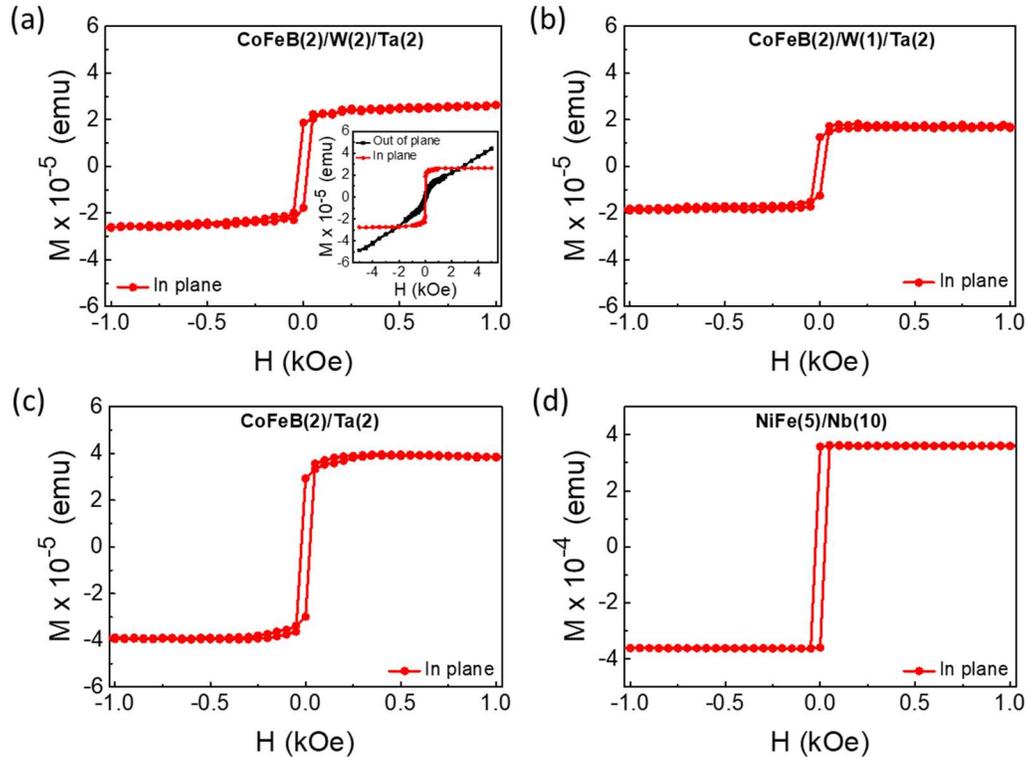

**Figure S3.** Magnetic-hysteresis loops from in-plane magnetic measurements using a vibrating sample magnetometer for (a) CoFeB(2)/W(2)/Ta(2), (b) CoFeB(2)/W(1)/Ta(2), (c) CoFeB(2)/Ta(2), and NiFe(5)/Nb(10) spintronic heterostructures. The numbers inside the small parentheses represent thicknesses or corresponding films in nm. Inset in (a) shows an out-of-plane M-H measurement for a comparison with the in-plane M-H measurement on the CoFeB(2)/W(2)/Ta(2) sample.

Vibrating sample magnetometer magnetic-hysteresis (M-H) measurements were performed using a quantum design magnetic properties measurements system. Figures S3(a) to (d) show the in-plane M-H loops of CoFeB(2)/W(2)/Ta(2), CoFeB(2)/W(1)/Ta(2), CoFeB(2)/Ta(2), and NiFe(5)/Nb(10) heterostructures, respectively. The number inside small parenthesis represents respective film thickness in the nm. Although a clear in-plane magnetic anisotropy is evident for all the samples but we have also performed the out-of-plane measurement for a few for completement and comparison. The in-plane and out-of-plane M-H loops for one of the samples, i.e., CoFeB(2)/W(2)/Ta(2) are shown in the inset of Fig. S3(a). The nonsaturating behavior in the case of out-of-plane M-H in the given range of field confirms the absence of any out-of-plane anisotropy. Usually, a perpendicular magnetic anisotropy occurs in the case of W-interlayer having high resistivity[5]. However, we do not find any change due to the W-insertion layer in M-H response of the CoFeB/W/Ta sample.

## S4. M-T measurements

We performed field-cooled (FC) and zero field-cooled (ZFC) magnetization measurements as a function of the sample temperature after cooling the heterostructures from 300 K to 10 K in an applied in-plane magnetic field of 600 (100) Oe (FC) and 0 Oe (ZFC). The 600 Oe and 100 Oe FC values are corresponding to the CoFeB/W/Ta and NiFe/Nb samples, respectively. FC and ZFC data were also collected while warming the heterostructures in a field of 600 Oe and 100 Oe. Figure S4 summarizes the above results obtained for the saturation magnetization ($M_S$) as a function of the sample temperature. It is evident from the figures that the temperature-dependent behaviour of the magnetization remains the same throughout the temperature range. Therefore, the influence of any temperature-dependent change of the magnetic properties can be safely avoided in our experiments for THz emission from them.

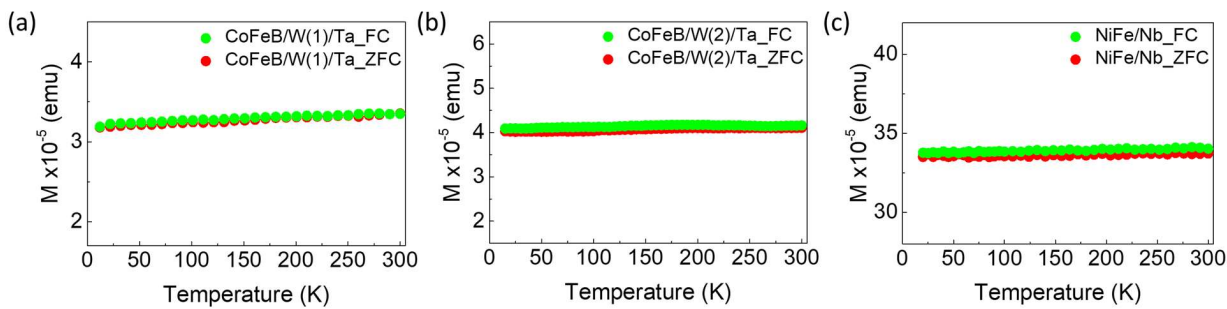

**Figure S4.** Saturation magnetization ($M_S$) as a function of varying sample temperature for (a) CoFeB/W(1)/Ta, (b) CoFeB/W(2)/Ta, and (c) NiFe/Nb samples obtained during field cooling and zero field cooling measurements.

## S5. Temperature-dependent THz time-domain spectroscopy setup

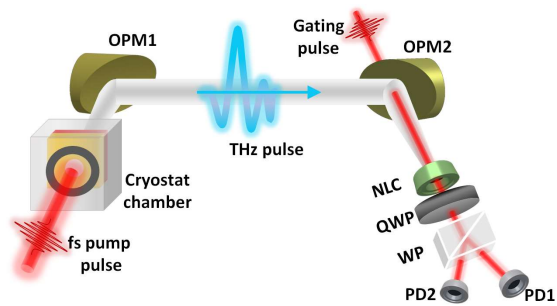

**Figure S5.** Cryogenically combined ultrafast optical-THz setup for temperature-dependent THz emission time-domain measurements. OPM: Off-axis parabolic mirror, QWP: Quarter-wave plate, NLC: Nonlinear crystal, WP: Wollaston prism, PD: Photodiode.

## S6. THz spectral bandwidth

Figure S6 presents typical fast-Fourier transformation (FFT) spectra of the broadband THz pulses emitted from our spintronic heterostrutures, here shown for the NiFe/Nb sample for its two different optical excitation configurations. In the first configuration, the sample is excited from the substrate side, while in the other, the optical excitation is done from the film sample side.

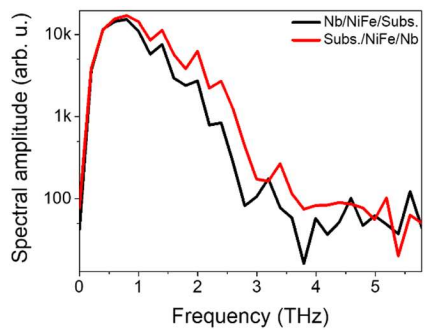

**Figure S6.** Typical Fast Fourier Transform (FFT) spectra of the THz emission time-domain signals generated from NiFe/Nb bilayer by exciting with the femtosecond pulase at 800 nm either from the substrate or the opposite side.

## S7. Temperature-dependent longitudinal electrical resistivity measurements

In Figure S7, we have presented results for temperature-dependent longitudinal resistivity by using four-point van der Pauw method for different material layers and their combinations/heterostructures. Parallel resistive model[1,6] was used to determine the resistivity of the W/Ta bilayer as shown in Figure S7(c).

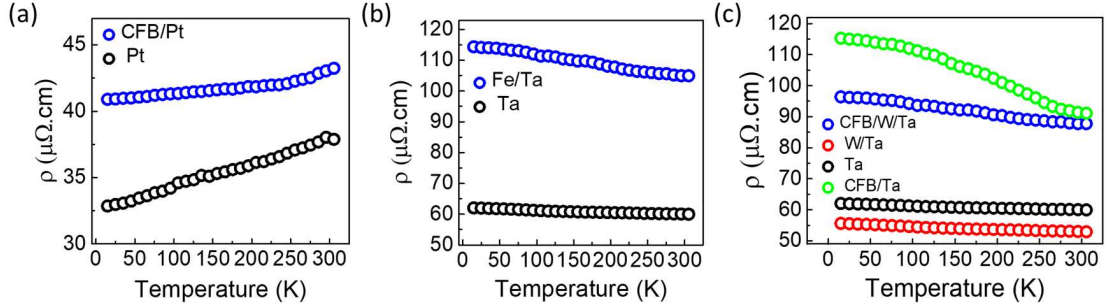

**Figure S7.** Temperature-dependent longitudinal resistivity ($\rho$ or $\rho_{xx}$) measured by four-point van der Pauw method for (a) CoFeB/Pt and Pt, (b) Fe/Ta and Ta, (c) CoFeB/W/Ta, CoFeB/Ta, W/Ta, and Ta thin film samples.

## S8. ISHE mediated THz emission from CoFeB/Ta bilayer

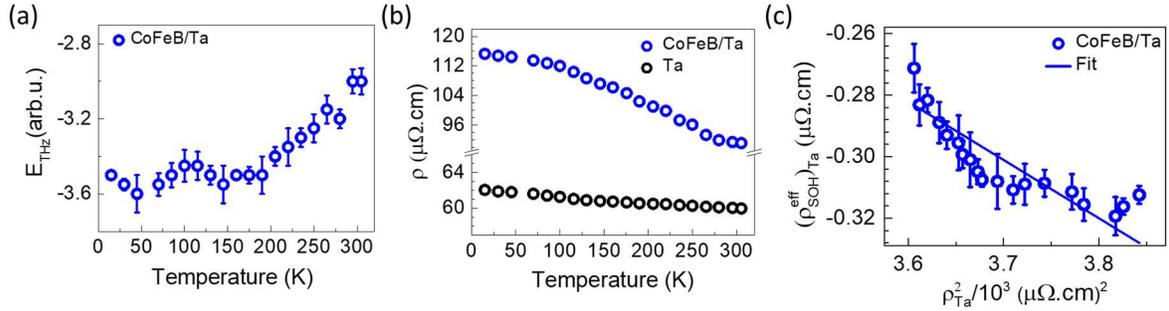

**Figure S8.** (a) Peak-to-peak THz signal variation with the varying sample temperature for CoFeB/Ta heterostructure. (b) Longitudinal resistivities ($\rho$) as a function of temperature measured by the four-point van der Pauw method. (c) Effective spin-orbital Hall resistivity vs squared longitudinal resistivity plot for Ta.

Temperature dependent behaviors of the emitted THz signal from CoFeB/Ta bilayer and its longitudinal resistivity are presented in Figs. S8(a) and S8(b), respectively. Using these resuls in Figs. S8(a) and (b), we determined the temperature-dependent behavior of the effective spin-orbital Hall resistivity vs squared longitudinal resistivity of Ta as shown in Fig. S8(c). Solid line in Fig. S8(c) is a linear fit to the data using Eq. (4) of the main manuscript. We note that Ta possesses a negatively valued large spin Hall conductivity and a smaller positively valued orbital Hall conductivity. The negative slope of the fitted curved in Fig. S8(c) and its value is consistent with the negative value of spin Hall conductivity in Ta signifying that ISHE is the source of THz radiation from CoFeB/Ta bilayer.

## S9. Orbital transport in CoFeB/W(1)/Ta

We also performed temperature-dependent THz emission and subsequent extraction of the effective intrinsic spin-orbit Hall conductivity of CoFeB/W/Ta trilayer having 1 nm thick W-insertion layer. Figure S9(a) shows the peak-to-peak amplitude of THz time-domain signal vs the sample temperature. The longitudinal resistivities of CoFeB/W(1)/Ta and W/Ta as a function of temperature are provided in Fig. S9(b). The information from Figs. S9(a) and S9(b) are used to extract the results in Fig. S9(c), where the effective spin-orbital Hall resistivity $\rho_{SH}^{eff}$ has been plotted as a function of squared longitudinal resistivity of W/Ta. In this figure, solid curve is linear fit to the data using Eq. 4 of the main manuscript. From the negative slope of the linear fit, we have determined the intrinsic spin Hall conductivity value of $(\sigma_{SH}^{int})^{eff} = -1250\ (\hbar/e)\ \Omega^{-1}cm^{-1}$.

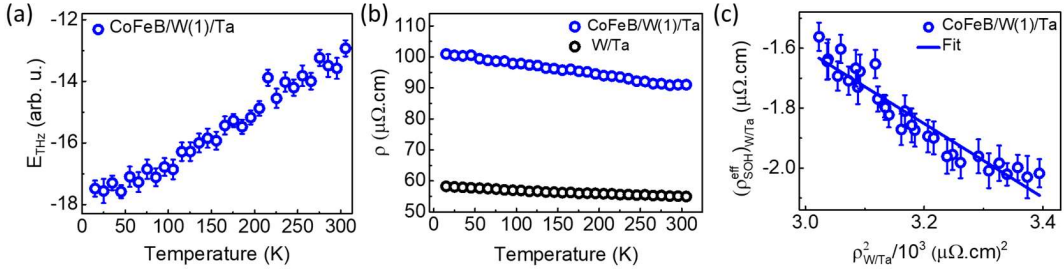

**Figure S9.** (a) Peak-to-peak THz signal amplitude variation with respect to the sample temperature for CoFeB/W(1)/Ta. (b) Temperature-dependent longitudinal resistivity of CoFeB/W/Ta and W/Ta samples. (c) Effective spin-orbital Hall resistivity $\rho_{SH}^{eff}$ as a function of the squared longitudinal resistivity of W/Ta.

## S10. Effective spin-orbital Hall resistivity from the THz electric field strength

The THz electric field strength ($E_{THz}$) from the measured electro-optic signal can be estimated using the standard procedure[1, 7, 8] that is discussed briefly here. Considering the fact that the spintronic emitters share almost a similar emission efficiency,[9] the THz electric field from the measured differential intensity ($\Delta I/I$) by electro-optic sampling in the ZnTe nonlinear crystal or our expeirmental setup, is estimated using the following expression.

$$E_{THz}\left(\frac{V}{cm}\right) = \left(\frac{\Delta I}{I}\right)\frac{2c}{\omega n^3 r_{41} L}\cdot\left(\frac{1}{\cos\alpha\ \sin 2\beta + 2\sin\ \cos 2\beta}\right) \quad (S1),$$

where, the c = 3x10⁸ m/s, ω = 2πν = 2πc/800nm, L = 0.5mm, n = 2.85[10], $r_{41}$ = 3.9x10⁻¹² m/V[7], α = 90⁰ and β = 180⁰ in our experiments for the THz electric field at the ZnTe crystal.

For the THz emission by ISHE in FM/NM-based heterostructures, the magnitude of the corresponding THz electric field depends on various parameters via the relation,[11, 12]

$$E(\omega)_{THz} = \frac{e.J_S(\omega).\theta_{SH}.\lambda_S}{(1+n)/Z_0 + \int_0^d \sigma(\omega,z)dz} \quad (S2)$$

Here, $n$, $Z_0$, $\theta_{SH}$, $e$, $J_S$, $\lambda_S$ and $d$ are substrate refractive index, vacuum impedance, spin Hall angle, elementary charge, spin current density, spin relaxation length and thickness of the heterostructure, respectively. The electronic charge is included in the above formula because $J_S$ and $J_C$ (charge current density) are used in the units of $\hbar$ and $e$, respectively. The shunt conductance $(1+n)/Z_0$ is much lower than conductance of the heterostructures $\int_0^d \sigma(\omega, z)dz$, therefore the former can be neglected in the estimation. Using $\int_0^d \sigma(\omega, z)dz = d/\rho_{FM/NM}$, Eq. (S2) changes to

$$E(\omega)_{THz} = \frac{e.J_S(\omega).\theta_{SH}.\lambda_S}{(d/\rho_{FM/NM})} \quad (S3),$$

where, a fact has been used that the conductivity depends only weakly on the THz frequencies[11, 13] and hence can be approximated to a constant. Ignoring any frequency-dependence of the other quantities, the corresponding relation in the time-domain becomes,

$$E(t)_{THz} = \frac{e.J_S(t).\theta_{SH}.\lambda_S}{(d/\rho_{FM/NM})} \quad (S4)$$

Therefore, the instantaneous value of the spin-current generated by ultrafast excitation can be quantified directly from the maximum of the time-domain THz signal. Also, the spin current is generated from FM layer is nearly temperature insensitive since all the measurements were performed well below the Curie temperature, as confirmed by M-T measurements shown in Section S4. Without loss of generality, the peak-to-peak value of the experimentally measured THz field has been considered for the latter in all our analysis. The spin Hall angle $\theta_{SH}$ is usually defined[14] as $\theta_{SH} = \rho_{SH}^{NM}/\rho_{xx}^{NM}$, where, $\rho_{SH}^{NM}$ and $\rho_{xx}^{NM}$ represent the spin Hall resistivity and longitudinal resistivity, respectively, of the NM layer. With all the above, a relationship between the spin Hall resistivity and the maximum of the measured THz field can be obtained as

$$\rho_{SH}^{NM} = E_{THz}\left(\frac{\rho_{xx}^{NM}}{\rho_{FM/NM}}\right)\left(\frac{d}{\lambda_S}\right)\frac{1}{e.J_S} \quad (S5)$$

For the case of THz emission by IOHE and ISHE, where both the spin current to charge current and orbital current to charge current conversions take place simultaneously, the above relation needs to be modified. IOHE becomes relevant if spin-orbit conversion takes place in the material layer via the spin-orbit correlation factor, $\eta_{L-S}$. Hence, for this case, the above relation gets modified to the following one,

$$E_{THz} = \frac{e.J_S.\lambda_{LS}}{(d/\rho_{FM/NM})}\left(\frac{\rho_{SH}^{NM}}{\rho_{xx}^{NM}} + \frac{\rho_{OH}^{NM}}{\rho_{xx}^{NM}}.\eta_{L-S}\right) \quad (S6)$$

Notice that, $J_S$ is retained in the above relation because it represents the maximum instantaneous spin current produced in the FM layer by the ultrafast optical excitation, however, the spin relaxation length has been substituted by the effective spin-orbit diffusion length,[15] $\lambda_{LS} = \sqrt{\lambda_L \lambda_S}$ ($\lambda_L$ and $\lambda_S$ being the orbital and spin diffusion lengths) and $\theta_{SH}$ has been substituted by an effective spin-orbit Hall angle[16], $\theta_{SOH}^{eff} = \left(\frac{\rho_{SH}^{NM}}{\rho_{xx}^{NM}} + \frac{\rho_{OH}^{NM}}{\rho_{xx}^{NM}}.\eta_{L-S}\right)$ to take into account the orbital Hall resistivity, $\rho_{OH}^{NM}$ of the NM layer. The $\theta_{SOH}^{eff}$ effectively takes care of the overall effect of spin and orbital currents, and their interconversion on the ultimate charge current and hence THz radiation produced.[15, 16] Equation (S6) is strictly valid for unit interfacial transparency,[16, 17] which can be rearranged to obtain a relation between the effective spin-orbit Hall resistivity, $\rho_{SOH}^{eff}$ and the measured THz signal as

$$\rho_{SOH}^{eff} = (\rho_{SH}^{NM} + \rho_{OH}^{NM}.\eta_{L-S}) = E_{THz}\left(\frac{\rho_{xx}^{NM}}{\rho_{FM/NM}}\right)\left(\frac{d}{\lambda_{LS}}\right)\frac{1}{e.J_S} \quad (S7)$$

Now, at each sample temperature $T$, the value of the experimentally measured THz signal $E_{THz}(T)$ in V/cm from Eq. (S1) can be used to calculate the effective spin-orbit Hall resistivity $\rho_{SOH}^{eff}(T)$ in the units of $\mu\Omega.cm$ using Eq. (S7) for the complete temperature-dependent analysis.

## S11. THz amplitude dependency on the ultrafast excitation fluence

Ultrafast excitation fluence dependence of the THz emission efficiecy of various heterostructures at the room tempreature was also evaluaed. Some of the representative results are presented in Fig. S10. It can be seen that the THz signal magnitude follows nearly linear dependence on the excitation fluences up to a large fluence value for all the samples. However, a slight deviation from the linear behaviour is observed for the NiFe/Nb and CoFeB/W/Ta heterostructures. We note that both types of these heterostructures emit THz radiation due to IOHE in the NM layer. A similar weakly saturating behaviour with respect to the excitation fluence has also been reported very recently by another group[18] in which case also, the THz radiation generation was majorly attributed to ultrafast IOHE in the NM layer.

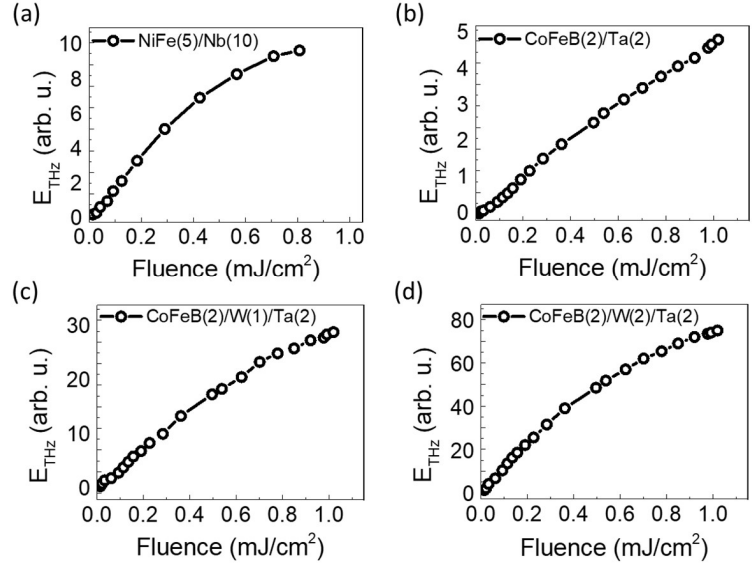

**Figure S10.** Peak to peak THz amplitude with the varying ultrafast pump fluence for (a) NiFe(5)/Nb(10), (b) CoFeB(2)/Ta(2), (c) CoFeB(2)/W(1)/Ta(2), and (d) CoFeB(2)/W(2)/Ta(2) heterostructures. The integers inside small parentheses represent thickness of respective layers in nm.

## S12. THz emission from the FM layer alone

We have also recorded a weak THz emission from the FM layers alone. The corresponding results for the NiFe and CoFeB are presented in Fig. S11. Figure S11(a) shows the THz time-doamin signal emitted from bare $Ni_{0.9}Fe_{0.1}$ sample. As evident from the figure, the THz signal polarity remains the same irrespective of the sample excitation geometry. The terms, back side and front side in the figure are designated for the substrate side and film side optical pumping, respectively. The THz emission from a different FM material layers has also been demonstrated in various reports in the literature.[19-22] In all such cases, the origin for the weak THz emission can be attributed mainly to either the anomalous Hall effect (AHE)[19, 21, 22] or ultrafast demagnetization[23-25] (UDM). Both of these mechanisms have different origins and follow different characteristics, which can be distinguished qualitatively by analyzing THz polarity behavior with respect to the sample excitation geometry. The polarity of the THz waveform is reversed when the sample is flipped, a phenomenon that can be attributed to the AHE[19, 21, 22] resulting from a change in the direction of the net backflow current. However, in the case of ultrafast demagnetization[23-25] (UDM), the THz waveform polarities remain the same upon sample flipping due to the fact that the magnetization dynamics is insensitive to the sample excitation direction geometry. Our results on NiFe(5) are found to be well aligned with the latter scenario as apparent from Fig. S11(a). We have compared the THz signals from bare NiFe and the NiFe/Nb bilayer in Figure S11(b). A significantly large difference between the THz signal magnitudes from the two sample clearly indicates only weak THz emission due to UDM in the NiFe layer. Furthermore, the consistency of our observation with the UDM in NiFe rules out any possibilities for THz emission from NiFe(5nm)/Nb(10nm) sample due to other mechanisms involving AHE[19, 21, 22] and structural inversion symmetry.[26] Similar outcomes are also observed in the case of bare CoFeB sample (see Fig. S11(c)) to reiterate that very weak THz emission from CoFeB layer is not due to AHE.

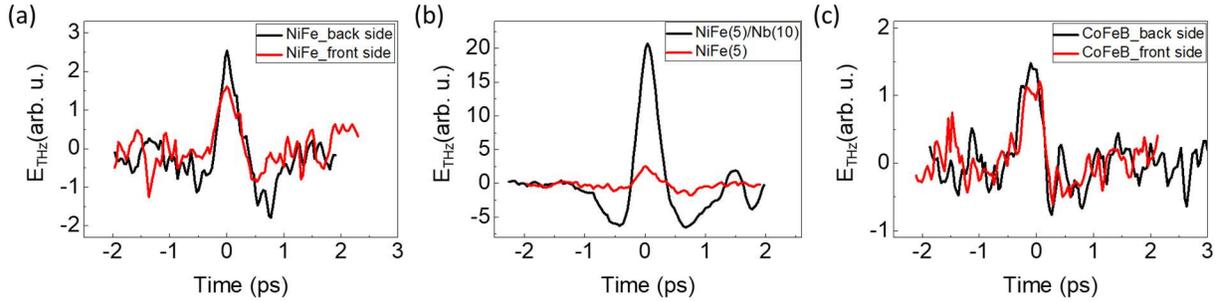

**Figure S11.** Time-domain THz waveforms generated from (a) bare NiFe layer under front- and back-side optical pumping, (b) NiFe and NiFe/Nb samples, and (c) bare CoFeB layer under front- and back-side optical pumping. The integers inside small parentheses are thicknesses of the respective layers in nm.

## S13. Nb thickness dependent THz emission from NiFe/Nb bilayer

Figure S12 presents results for the THz emission from NiFe/Nb samples and to compare the THz signal magnitude with respect to the thickness of the Nb layer. Figure S12(a) presents the raw data and explicit Nb-thickness dependence of the THz signal magnitude that is normalized with both optical absorbance and THz impedance is shown in Figure S12(b). The latter is calculated[18, 27, 28] by using transmitted THz signal amplitudes from the samples and reference (see Fig. S12 (c)), which are obtained by performing the time-domain THz transmission measurements with an ultrafast air-plasma based THz sources. For these measurements on NiFe/Nb bilayers of varying Nb thickness, the NiFe layer thickness was kept the same at 5 nm for all. We can see that the THz signal polarity is same for all the samples and its magnitude decreases strongly for higher thicknesses of the Nb-layer in NiFe/Nb bilayer. As brought out clearly in our main manuscript that IOHE is the source of THz radiation from the NiFe/Nb bilayer, we believe that our results on the Nb-thickness dependence of the THz emission (Fig. S12(b)) is a new value addition to the field. From Fig. S12(b), we also conclude that the orbital current diffusion length in Nb is of ~25 nm, that is in the same order as the spin current diffusion length in it at the room temperature.[29] A similar value persists at the low temperatures as well.

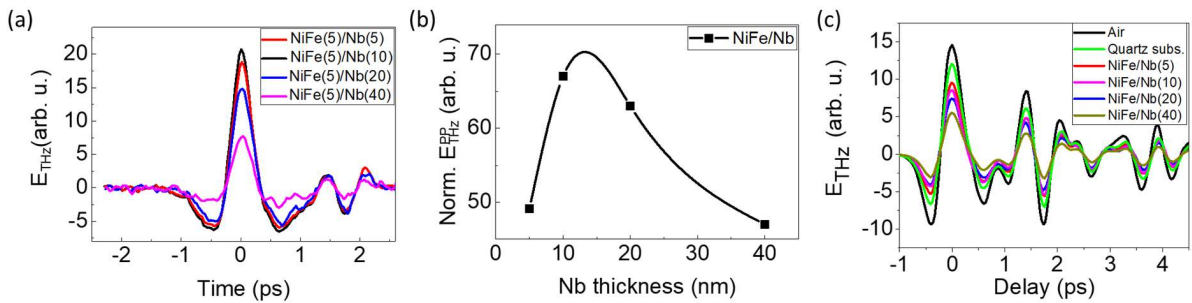

**Figure S12.** (a) Time-domain THz waveforms emitted from NiFe/Nb bilayer with the varying thickness of Nb layer. (b) Corresponding peak-to-peak THz signal amplitude variation with the Nb layer thcikness. (c) As recorded THz transmission signals from different NiFe/Nb samples of varying Nb thickness, including the reference signal in air and through the quartz substrate. The THz source for the measurements in (c) was a dual color ultrafast air plasma.

## S14. THz emission from the CoFeB/W bilayer

Figure S13(a) presents results from the time-domain THz emission measurements on the CoFeB/W bilayers. A comparison in the THz signal generation from CoFeB(2)/W(2)/Ta(2), CoFeB(2)/W(2), and CoFeB(2)/Ta(2) samples is provided in Fig. S13(b). The number inside the small parenthesis is the respective film thickness in nm. From these results, certain conclusions can be made: (i) THz signal emitted from CoFeB/W(2nm) bilayer sample is ~2.5 times higher than that from CoFeB/Ta(2nm) whereas their polarities are same. (ii) ISHE is the origin of THz emission from both the CoFeB/W(2nm) and CoFeB/Ta(2nm) bilayers. Analogous to CoFeB/Ta(2nm) bilayer, ISHE is the origin of THz emission from CoFeB/W(2nm). The similar polarity and different THz amplitudes are quite consistent with the sign and magnitude of the spin Hall angles in W and Ta. (iii) THz signal from CoFeB/W bilayer decreases if the W layer thickness is increased from 2 to 3 nm. This clearly indicates that the spin diffusion length in the W-layer, and hence it's optimum thickness for the THz emission via ISHE, is about 2-3 nm. Such a value of the spin diffusion length in W-layer matches well with the literature.[30] (iv) CoFeB/W(2)/Ta(2) trilayer sample emits ~10 times stronger signal than from CoFeB/Ta(2) bilayer and ~4 times higher THz signal as compared to that from the CoFeB/W(2) sample. For such a large difference in the THz emission efficiency of the trilayer as against the bilayer counterparts is argued to originate from W-insertion layer mediated enhanced orbital transport and its conversion to charge current in the Ta layer of the trilayer heterostructure.

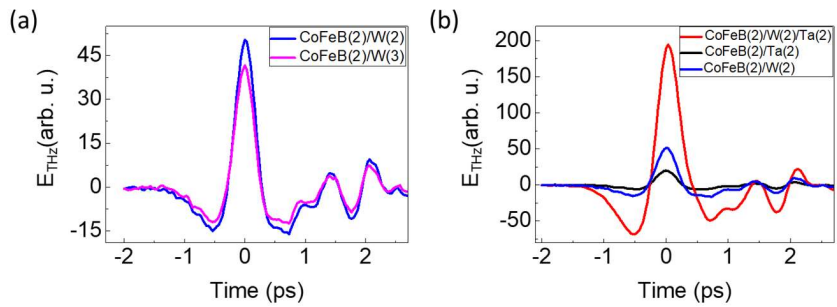

**Figure S13.** (a) Time-domain THz waveforms emitted from CoFeB(2)/W(2) and CoFeB(2)/W(3) bilayers. (b) Comparisons of THz signal amplitude for CoFeB(2)/W(2)/Ta(2), CoFeB(2)/W(2), and CoFeB(2)/Ta(2) samples. The numbers inside small parentheses represent thickness of the respective layer in nm.

## S15. Temperature-dependent optical transmission of the substrate

Since we have performed all the THz emission measurements in the transmission mode, where the optical beam traverses the substrate first before reaching the film, it becomes important to check any temperature-dependent change in the optical transmission from the quartz substrate. Therefore, we have carried out optical transmission measurements on our bare quartz substrate under similar conditions of pump fluence and temperatures used for samples in our study. The corresponding result is shown in Fig. S14. There is hardly any change in the transmission, and it is less than a percent while traversing from the lowest temperature to the room temperature. We measured a negligible (<2%) change in the THz emission efficiency from our samples corresponding to the variation in the excitation power mentioned above. Such an insignificant excitation power variation does not affect the large temperature dependence of the THz signal as measured from our samples.

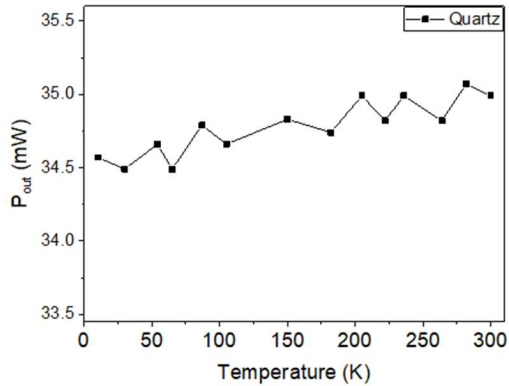

**Figure S14.** Temperature-dependent optical transmission at 800 nm from the substrate.

## S16. One-dimensional spin-orbit drift diffusion model for analyzing the enhancement in the orbital current with the thickness of the W-insertion layer

As shown in Fig. 4 (c) of our manuscript, the THz signal from CoFeB/W(t)/Ta trilayer heterostructure increases with the increasing thickness (t) of the W-insertion layer. This observation is consistent with the fact that heavy metal W possesses large negative valued spin-orbit correlation factor owing to which efficient spin-orbit conversion occurs in it. Long diffusion length for the orbital current and high orbital Hall conductivity in the adjacent Ta layer profuse in tandem to generate stronger THz signal via IOHE from CoFeB/W/Ta trilayer and efficiency increases with the varying thickness of W-insertion layer. Following the assertions of spin-orbit interconversions by Sala et al.,[15] and the associated coupled differential equations for the chemical potentials and current densities related to the spin and orbital degrees, the thickness dependent enhancement in the OHE or the ultrafast IOHE in our case can be analyzed. Figure S15 presents the response of the orbital current as a function of the W-insertion layer calculated using the phenomenological model of Sala et al.[15] for the orbital current given by the relation,

$$J_L(z_{NM}) = -\left(\frac{\sigma_S \mp \frac{\sigma_L}{\lambda_{LS}^2 \gamma_2}}{1 - \frac{\gamma_2}{\gamma_1}}\right)\frac{E}{2}\text{Sech}^2\left(\frac{z_{NM}}{2\lambda_1}\right) - \left(\frac{\sigma_S \mp \frac{\sigma_L}{\lambda_{LS}^2 \gamma_1}}{1 - \frac{\gamma_1}{\gamma_2}}\right)\frac{E}{2}\text{Sech}^2\left(\frac{z_{NM}}{2\lambda_2}\right) + \sigma_L E \qquad (S8)$$

Here, different parameters have meaning as given in the original paper,[15] E is the applied external field in typical OHE settings, $z_{NM}$ is the thickness of the heavy metal layer and so on. In generating the qualitative result of Fig. S15, we have used the fact that $\sigma_S < 0$ and $\sigma_L > 0$ for the heavy metal W.[31, 32] Clearly, a larger thickness of the W-insertion layer supports larger spin-orbital conversion. Hence, stronger orbital current is injected into the adjacent Ta layer which possesses much stronger orbital Hall conductivity than the spin Hall conductivity, thereby, resulting into much stronger orbit-charge conversion via IOHE and hence stronger THz emission from the trilayer with thicker W-insertion layer. We believe that our extensive temperature-dependent experiments for the THz emission via IOHE in CoFeB/W/Ta have much scope for further theoretical exploration and more experiments studies on such systems in future.

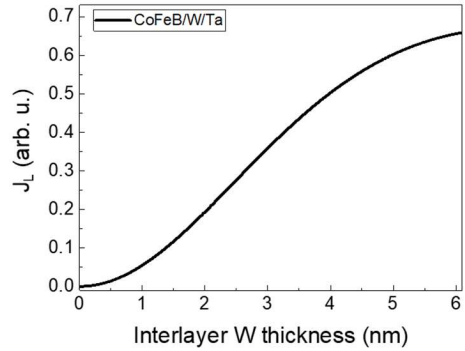

**Figure S15.** Increase in the orbital current with the increasing thickness of W-insertion layer in CoFeB/W/Ta trilayer.

## References

1. Kumar, S.; Kumar, S., Large interfacial contribution to ultrafast THz emission by inverse spin Hall effect in CoFeB/Ta heterostructure. *iScience* **2022**, *25* (8), 104718.
2. Kumar, S.; Nivedan, A.; Singh, A.; Kumar, Y.; Malhotra, P.; Tondusson, M.; Freysz, E.; Kumar, S., Optical damage limit of efficient spintronic THz emitters. *iScience* **2021**, *24* (10), 103152.
3. Jhajhria, D.; Behera, N.; Pandya, D. K.; Chaudhary, S., Dependence of spin pumping in W/CoFeB heterostructures on the structural phase of tungsten. *Physical Review B* **2019**, *99* (1), 014430.
4. Kumar, A.; Bansal, R.; Chaudhary, S.; Muduli, P. K., Large spin current generation by the spin Hall effect in mixed crystalline phase Ta thin films. *Physical Review B* **2018**, *98* (10), 104403.
5. Li, S. K.; Zhao, X. T.; Liu, W.; Wang, T. T.; Zhao, X. G.; Zhang, Z. D., Enhanced spin-orbit torques and perpendicular magnetic anisotropy in CoFeB/MgO structures with Ta/W bilayer. *AIP Advances* **2018**, *8* (6), 065007.
6. Chen, Y.-Y.; Juang, J.-Y., Finite element analysis and equivalent parallel-resistance model for conductive multilayer thin films. *Measurement Science and Technology* **2016**, *27* (7), 074006.
7. Planken, P. C. M.; Nienhuys, H.-K.; Bakker, H. J.; Wenckebach, T., Measurement and calculation of the orientation dependence of terahertz pulse detection in ZnTe. *J. Opt. Soc. Am. B* **2001**, *18* (3), 313-317.

8. Cheng, L.; Wang, X.; Yang, W.; Chai, J.; Yang, M.; Chen, M.; Wu, Y.; Chen, X.; Chi, D.; Goh, K. E. J.; Zhu, J.-X.; Sun, H.; Wang, S.; Song, J. C. W.; Battiato, M.; Yang, H.; Chia, E. E. M., Far out-of-equilibrium spin populations trigger giant spin injection into atomically thin MoS2. *Nature Physics* **2019,** *15* (4), 347-351.
9. Yang, D.; Liang, J.; Zhou, C.; Sun, L.; Zheng, R.; Luo, S.; Wu, Y.; Qi, J., Powerful and Tunable THz Emitters Based on the Fe/Pt Magnetic Heterostructure. *Advanced Optical Materials* **2016,** *4* (12), 1944-1949.
10. Li, H. H., Refractive Index of ZnS, ZnSe, and ZnTe and Its Wavelength and Temperature Derivatives. *Journal of Physical and Chemical Reference Data* **1984,** *13* (1), 103-150.
11. Seifert, T. S.; Tran, N. M.; Gueckstock, O.; Rouzegar, S. M.; Nadvornik, L.; Jaiswal, S.; Jakob, G.; Temnov, V. V.; Münzenberg, M.; Wolf, M.; Kläui, M.; Kampfrath, T., Terahertz spectroscopy for all-optical spintronic characterization of the spin-Hall-effect metals Pt, W and Cu80Ir20. *Journal of Physics D: Applied Physics* **2018,** *51* (36), 364003.
12. Seifert, T.; Jaiswal, S.; Martens, U.; Hannegan, J.; Braun, L.; Maldonado, P.; Freimuth, F.; Kronenberg, A.; Henrizi, J.; Radu, I.; Beaurepaire, E.; Mokrousov, Y.; Oppeneer, P. M.; Jourdan, M.; Jakob, G.; Turchinovich, D.; Hayden, L. M.; Wolf, M.; Münzenberg, M.; Kläui, M.; Kampfrath, T., Efficient metallic spintronic emitters of ultrabroadband terahertz radiation. *Nature Photonics* **2016,** *10*, 483.
13. Seifert, T.; Martens, U.; Günther, S.; Schoen, M. A. W.; Radu, F.; Chen, X. Z.; Lucas, I.; Ramos, R.; Aguirre, M. H.; Algarabel, P. A.; Anadón, A.; Körner, H. S.; Walowski, J.; Back, C.; Ibarra, M. R.; Morellón, L.; Saitoh, E.; Wolf, M.; Song, C.; Uchida, K.; Münzenberg, M.; Radu, I.; Kampfrath, T., Terahertz Spin Currents and Inverse Spin Hall Effect in Thin-Film Heterostructures Containing Complex Magnetic Compounds. *SPIN* **2017,** *07* (03), 1740010.
14. Sagasta, E.; Omori, Y.; Isasa, M.; Gradhand, M.; Hueso, L. E.; Niimi, Y.; Otani, Y.; Casanova, F., Tuning the spin Hall effect of Pt from the moderately dirty to the superclean regime. *Physical Review B* **2016,** *94* (6), 060412.
15. Sala, G.; Gambardella, P., Giant orbital Hall effect and orbital-to-spin conversion in 3d, 5d, and 4f metallic heterostructures. *Physical Review Research* **2022,** *4* (3), 033037.
16. Lee, S.; Kang, M.-G.; Go, D.; Kim, D.; Kang, J.-H.; Lee, T.; Lee, G.-H.; Kang, J.; Lee, N. J.; Mokrousov, Y.; Kim, S.; Kim, K.-J.; Lee, K.-J.; Park, B.-G., Efficient conversion of orbital Hall current to spin current for spin-orbit torque switching. *Communications Physics* **2021,** *4* (1), 234.
17. Go, D.; Freimuth, F.; Hanke, J.-P.; Xue, F.; Gomonay, O.; Lee, K.-J.; Blügel, S.; Haney, P. M.; Lee, H.-W.; Mokrousov, Y., Theory of current-induced angular momentum transfer dynamics in spin-orbit coupled systems. *Physical Review Research* **2020,** *2* (3), 033401.
18. Seifert, T. S.; Go, D.; Hayashi, H.; Rouzegar, R.; Freimuth, F.; Ando, K.; Mokrousov, Y.; Kampfrath, T., Time-domain observation of ballistic orbital-angular-momentum currents with giant relaxation length in tungsten. *Nature Nanotechnology* **2023**.
19. Zhang, Q.; Luo, Z.; Li, H.; Yang, Y.; Zhang, X.; Wu, Y., Terahertz Emission from Anomalous Hall Effect in a Single-Layer Ferromagnet. *Physical Review Applied* **2019,** *12* (5), 054027.
20. Huang, L.; Lee, S.-H.; Kim, S.-D.; Shim, J.-H.; Shin, H. J.; Kim, S.; Park, J.; Park, S.-Y.; Choi, Y. S.; Kim, H.-J.; Hong, J.-I.; Kim, D. E.; Kim, D.-H., Universal field-tunable terahertz emission by ultrafast photoinduced demagnetization in Fe, Ni, and Co ferromagnetic films. *Scientific Reports* **2020,** *10* (1), 15843.
21. Liu, Y.; Cheng, H.; Xu, Y.; Vallobra, P.; Eimer, S.; Zhang, X.; Wu, X.; Nie, T.; Zhao, W., Separation of emission mechanisms in spintronic terahertz emitters. *Physical Review B* **2021,** *104* (6), 064419.
22. Mottamchetty, V.; Rani, P.; Brucas, R.; Rydberg, A.; Svedlindh, P.; Gupta, R., Direct evidence of terahertz emission arising from anomalous Hall effect. *Scientific Reports* **2023,** *13* (1), 5988.
23. Beaurepaire, E.; Turner, G. M.; Harrel, S. M.; Beard, M. C.; Bigot, J. Y.; Schmuttenmaer, C. A., Coherent terahertz emission from ferromagnetic films excited by femtosecond laser pulses. *Applied Physics Letters* **2004,** *84* (18), 3465-3467.
24. Beaurepaire, E.; Merle, J. C.; Daunois, A.; Bigot, J. Y., Ultrafast Spin Dynamics in Ferromagnetic Nickel. *Physical Review Letters* **1996,** *76* (22), 4250-4253.
25. Zhang, W.; Maldonado, P.; Jin, Z.; Seifert, T. S.; Arabski, J.; Schmerber, G.; Beaurepaire, E.; Bonn, M.; Kampfrath, T.; Oppeneer, P. M.; Turchinovich, D., Ultrafast terahertz magnetometry. *Nature Communications* **2020,** *11* (1), 4247.
26. Rouzegar, R.; Brandt, L.; Nádvorník, L.; Reiss, D. A.; Chekhov, A. L.; Gueckstock, O.; In, C.; Wolf, M.; Seifert, T. S.; Brouwer, P. W.; Woltersdorf, G.; Kampfrath, T., Laser-induced terahertz spin transport in magnetic nanostructures arises from the same force as ultrafast demagnetization. *Physical Review B* **2022,** *106* (14), 144427.
27. Wang, P.; Feng, Z.; Yang, Y.; Zhang, D.; Liu, Q.; Xu, Z.; Jia, Z.; Wu, Y.; Yu, G.; Xu, X.; Jiang, Y., Inverse orbital Hall effect and orbitronic terahertz emission observed in the materials with weak spin-orbit coupling. *npj Quantum Materials* **2023,** *8* (1), 28.
28. Zhang, H.; Feng, Z.; Zhang, J.; Bai, H.; Yang, H.; Cai, J.; Zhao, W.; Tan, W.; Hu, F.; Shen, B.; Sun, J., Laser pulse induced efficient terahertz emission from Co/Al heterostructures. *Physical Review B* **2020,** *102* (2), 024435.
29. Jeon, K.-R.; Ciccarelli, C.; Kurebayashi, H.; Wunderlich, J.; Cohen, L. F.; Komori, S.; Robinson, J. W. A.; Blamire, M. G., Spin-Pumping-Induced Inverse Spin Hall Effect in Nb/Ni80Fe20 Bilayers and its Strong Decay Across the Superconducting Transition Temperature. *Physical Review Applied* **2018,** *10* (1), 014029.

30. Wang, T.-C.; Chen, T.-Y.; Wu, C.-T.; Yen, H.-W.; Pai, C.-F., Comparative study on spin-orbit torque efficiencies from W/ferromagnetic and W/ferrimagnetic heterostructures. *Physical Review Materials* **2018,** *2* (1), 014403.
31. Hayashi, H.; Jo, D.; Go, D.; Gao, T.; Haku, S.; Mokrousov, Y.; Lee, H.-W.; Ando, K., Observation of long-range orbital transport and giant orbital torque. *Communications Physics* **2023,** *6* (1), 32.
32. Salemi, L.; Oppeneer, P. M., First-principles theory of intrinsic spin and orbital Hall and Nernst effects in metallic monoatomic crystals. *Physical Review Materials* **2022,** *6* (9), 095001.